\documentclass[8.5pt,twoside,twocolumn]{article}
\usepackage[super,sort&compress,comma]{natbib} 
\usepackage{mhchem}
\usepackage{times,mathptm}
\usepackage{sectsty}
\usepackage{balance} 
\usepackage{graphicx} 
\usepackage{dcolumn}
\usepackage{lastpage}
\usepackage[format=plain,justification=raggedright,singlelinecheck=false,font=small,labelfont=bf,labelsep=space]{caption} 
\usepackage{fancyhdr}
\begin{document}

\thispagestyle{plain}
\fancypagestyle{plain}{
\renewcommand{\headrulewidth}{1pt}}
\renewcommand{\thefootnote}{\fnsymbol{footnote}}
\renewcommand\footnoterule{\vspace*{1pt}%
\hrule width 3.4in height 0.4pt \vspace*{5pt}} 
\setcounter{secnumdepth}{5}

\makeatletter 
\def\subsubsection{\@startsection{subsubsection}{3}{10pt}{-1.25ex
plus -1ex minus -.1ex}{0ex plus 0ex}{\normalsize\bf}} 
\def\paragraph{\@startsection{paragraph}{4}{10pt}{-1.25ex plus
-1ex minus -.1ex}{0ex plus 0ex}{\normalsize\textit}} 
\renewcommand\@biblabel[1]{#1}            
\renewcommand\@makefntext[1]%
{\noindent\makebox[0pt][r]{\@thefnmark\,}#1}
\makeatother 
\renewcommand{\figurename}{\small{Fig.}~}
\sectionfont{\large}
\subsectionfont{\normalsize} 

\fancyfoot{}
\fancyfoot[RO]{\footnotesize{\sffamily{1--\pageref{LastPage}
~\textbar  \hspace{2pt}\thepage}}}
\fancyfoot[LE]{\footnotesize{\sffamily{\thepage~\textbar\hspace{3.45cm}
1--\pageref{LastPage}}}}
\fancyhead{}
\renewcommand{\headrulewidth}{1pt} 
\renewcommand{\footrulewidth}{1pt}
\setlength{\arrayrulewidth}{1pt}
\setlength{\columnsep}{6.5mm}
\setlength\bibsep{1pt}

\twocolumn[
  \begin{@twocolumnfalse}
\noindent\LARGE{\textbf{Spatial and temporal dynamical
heterogeneities approaching the binary colloidal glass transition}}
\vspace{0.6cm}

\noindent\large{\textbf{Takayuki Narumi,\textit{$^{a}$} 
Scott V. Franklin,\textit{$^{b}$} 
Kenneth W. Desmond,\textit{$^{c}$} 
Michio Tokuyama,\textit{$^{d}$} 
and
Eric R. Weeks$^{\ast}$\textit{$^{c}$}}}
\vspace{0.5cm}

\noindent \normalsize{
We study concentrated binary colloidal suspensions, a model system
which has a glass transition as the volume fraction $\phi$ of
particles is increased.  We use confocal microscopy to directly
observe particle motion within dense samples with $\phi$ ranging
from 0.4 to 0.7.  Our binary mixtures have a particle diameter
ratio $d_S/d_L=1/1.3$ and particle number ratio $N_S/N_L=1.56$,
which are chosen to inhibit crystallization and enable long-time
observations.  Near the glass transition we find that particle
dynamics are heterogeneous in both space and time.  The most mobile
particles occur in spatially localized groups.  The length scales
characterizing these mobile regions grow slightly as the glass
transition is approached, with the largest length scales seen
being $\sim 4$ small particle diameters.  We also study temporal
fluctuations using the dynamic susceptibility $\chi_4$, and find
that the fluctuations grow as the glass transition is approached.
Analysis of both spatial and temporal dynamical heterogeneity show
that the smaller species play an important role in facilitating
particle rearrangements.  The glass transition in our sample
occurs at $\phi_g \approx 0.58$, with characteristic signs of
aging observed for all samples with $\phi>\phi_g$.
}
\vspace{0.5cm}
 \end{@twocolumnfalse}
  ]




\footnotetext{\textit{$^{a}$~Department
of Applied Quantum Physics and Nuclear 
Engineering, Kyushu University, Fukuoka, Japan 819-0395.  E-mail:
narumi@athena.ap.kyushu-u.ac.jp}}
\footnotetext{\textit{$^{b}$~Department of Physics, 
Rochester Institute of Technology, Rochester, NY 14623-5603}}
\footnotetext{\textit{$^{c}$~Department of Physics, 
		Emory University, Atlanta, Georgia 30322}}
\footnotetext{\textit{$^{d}$~WPI Advanced Institute for Material
Research and Institute of Fluid Science,
Tohoku University, Sendai, Japan 980-8577}}

\section{Introduction}

As the temperature of a glass-forming liquid is lowered,
the viscosity rises by many orders of magnitude, becoming
experimentally difficult to measure, with little change in
the structure \cite{angell95,stillinger95,angell00}.
The origin of the slowing dynamics is not yet clear, despite much
prior work.
One intriguing observation is that as a sample approaches the
glass transition, the motion within the sample becomes spatially
heterogeneous \cite{gibbs65,ediger00}.  While overall motion within
the sample slows, some regions exhibit faster dynamics than the
rest, and over time these mobile regions appear and disappear
throughout the sample.  Particles within the mobile region move
cooperatively, forming spatially extended clusters and strings
\cite{donati98}.

One technique for studying the glass transition is the use of
colloidal suspensions \cite{pusey86}.  These are composed of small solid
particles suspended in a solvent.  The particles need to be small
enough to undergo Brownian motion, so particle diameters are
typically $10 - 5000$ nm.  The key control parameter is the
volume fraction $\phi$.  For a monodisperse sample (all particles
similar in size), the sample becomes glassy for $\phi > \phi_g
\approx 0.58$ \cite{pusey86,vanmegen91}.  The glass transition in
colloidal samples has been studied extensively by light scattering,
microscopy, and other techniques.  Colloidal samples exhibit many
behaviors seen in molecular glasses, such as 
dramatic increases in viscosity \cite{segre95,cheng02},
strongly slowing relaxation time scales
\cite{vanmegen91,mason95glass,bartsch95,vanmegen98,brambilla09,vanmegen10,brambilla10},
microscopic disorder \cite{vanblaaderen95},
spatially heterogeneous dynamics
\cite{marcus99,kegel00,weeks00,konig05}, 
aging behavior for glassy samples
\cite{vanmegen98,courtland03,cipelletti03,kegel04,cipelletti05,cianci06ssc,yunker09},
and sensitivity to finite size effects
\cite{nugent07prl,sarangapani08}.  
Light scattering allows careful study of the average behavior
of millions of colloidal particles, while microscopy techniques
observe the detailed behavior of a few thousand particles.
These complementary techniques have resulted in connections
between different aspects of glassy behavior:  for example,
showing that aging is temporally and spatially heterogeneous
\cite{courtland03,cipelletti03}, and connecting dynamical
heterogeneity with the slowing relaxation time scales
\cite{vanmegen91,vanmegen98,kegel00,weeks00,conrad06}.

In this paper, we study the glass transition of binary colloidal
suspensions using confocal microscopy.  We use binary suspensions
(mixtures of two particle sizes) to inhibit crystallization.
This allows us to take data over many hours, a time scale in
which a monodisperse sample would crystallize \cite{zhu97,gasser01}.
Furthermore, this lets us investigate the role the two particle
species play in the dynamics; prior work has suggested that
small particles play a lubricating role in the local dynamics
\cite{lynch08}.  Prior studies have also seen a connection
between the local structure and the mobility of particles
\cite{weeks02,conrad05,harrowell04,widmercooper05}.  Using a
binary sample results in more obvious structural variations due
to spatial variability of the composition, helping highlight how
structure influences the dynamics.

The confocal microscope enables direct visualization of the
interior of the sample, and we follow the motion of several thousand
colloidal particles within each sample \cite{prasad07}.  Particles
move in spatially heterogeneous groups, and we characterize
this motion using two-particle two-time correlation functions
\cite{poole98, donati99, doliwa00} that have previously been
used on monodisperse suspensions \cite{weeks07cor}.  From these
we extract a length scale for the heterogeneity, which increases
as the glass transition is approached.  By simultaneously tracking
both large and small particles, we can observe the similarities and
differences between the two species' dynamics.  In particular, we
see that small local composition fluctuations influence mobility.
As might be expected, regions with more large particles tend to
be less mobile, while those with more small particles tend to be
more mobile.  Additionally, we study temporal heterogeneity using
a different correlation function, the dynamic susceptibility
$\chi_4$
\cite{glotzer00_2,glotzer03a,keys07}, which has not been previously
applied to colloidal data.  As measured by this correlation
function, the temporal heterogeneity increases as the glass
transition is approached.

\section{Experimental Method}
\label{sec:experiment}

We prepare suspensions of poly-(methyl-methacrylate) (PMMA) colloids
stabilized sterically by a thin layer of poly-12-hydroxystearic
acid \cite{pusey86}.  We use a binary mixture
with a large particle
mean radius $a_{L}= 1.55$~$\mu $m and small particle mean radius
$a_{S} = 1.18$~$\mu$m, the same size particles used in a prior
study by our group \cite{nugent07prl}.  The polydispersity
is 5\%; each individual particle species can crystallize in a
monodisperse suspension.  Separately from the polydispersity, the
mean particle radii each have an uncertainty of $\pm 0.02$~$\mu$m.
The number ratio of small particles to large particles is $N_S /
N_L = 1.56$, resulting in a volume fraction ratio $\phi_S / \phi_L
\approx 0.70$.  The control parameter is the total volume fraction
$\phi=\phi_{S}+\phi_{L}$.  Crystallization and segregation were
not observed to occur during the course of our measurements.
All particles are fluorescently dyed and suspended in a density-
and index-matched mixture of decalin and cyclohexyl bromide
to prevent sedimentation and allow us to see into the sample.
Particles are slightly charged as a result of the dyeing process and
this particular solvent mixture \cite{gasser01}.  Nonetheless, we
use the hard sphere volume fraction $\phi$ as the control parameter.
The hard sphere radii ($a_L, a_S$) are determined from diffusion
measurements of the individual species at dilute concentrations
($\phi < 0.01$).

\begin{figure}[!ht]
\begin{center}
\includegraphics[width=7.5cm]{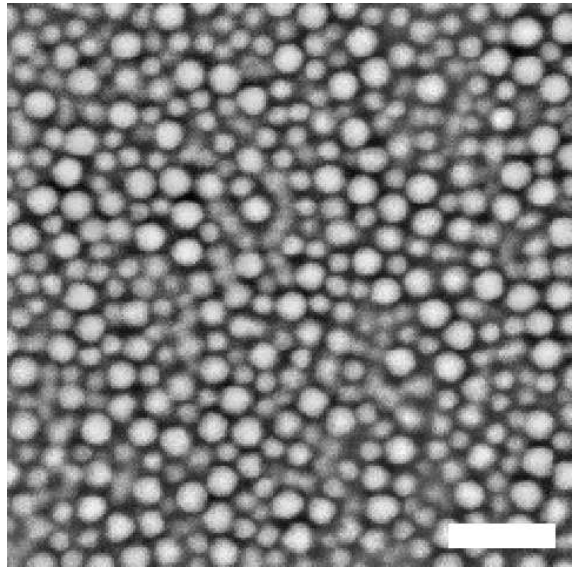}
\end{center}
\caption{A two-dimensional image of our sample taken by a confocal
microscope.  The scale bar represents 10 $\mu$m.}
\label{fig:confocal_image}
\end{figure}

Suspensions are sealed in microscope chambers and confocal
microscopy is used to observe the particle dynamics at ambient
temperature \cite{prasad07,dinsmore01}.  A representative
two-dimensional image is shown in Fig.~\ref{fig:confocal_image}.
A volume of $55 \times 55 \times 20 $~$\mu$m$^3$ can be taken
at speeds of up to 1 Hz.  (As will be shown later, in these
concentrated samples, particles do not move significantly on this
time scale.)  To avoid influences from the walls, we focus at
least 25~$\mu$m away from the coverslip.

Within each three-dimensional image, we identify both large
and small particles.  This is accomplished with a
single convolution that identifies spherical, bright regions
\cite{crocker96}; the convolution kernel is a three-dimensional
Gaussian with a width chosen to match the size of the image of
a large particle.  Each local maximum after the convolution is
identified as a particle \cite{crocker96}.
The distribution of particle brightnesses is bimodal
with little overlap, and so small and large particles
can be easily distinguished.  Our method is the
same as is often used to measure particle positions in two
dimensions, which normally achieves sub-pixel resolution in
particle positions \cite{crocker96}.  However, given that a single
convolution kernel is used to identify both particle types, when
applied to our binary samples we do not achieve
sub-pixel accuracy.  Instead, our uncertainty in locating particle
positions is linked to the pixel size and is 0.2~$\mu$m in $x$
and $y$, and 0.3~$\mu$m in $z$.  We do have accurate discrimination
between large and small particles with this method, with less than
1\% of the particles misidentified, checked by visual inspection.
For a few particles, it is hard to distinguish if they are large
or small (because they are small but unusually bright,
or large but unusually dim).  These particles are assigned
to the size that they appear to be the majority of the time.

After identifying the particle positions, they are tracked
using standard software \cite{crocker96,dinsmore01}.  The key
requirement is that particles move less between time steps than
their interparticle spacing, which is easily satisfied in our dense
glassy samples.  We take images once every 10 - 150~s, depending on
the volume fraction, in each case making sure that the acquisition
rate is sufficiently rapid to capture all particle movements.

\section{Results and Discussion} \label{sec:result}

\subsection{Structural characteristics}

\begin{figure}[!ht]
\begin{center}
\includegraphics[width=8.0cm]{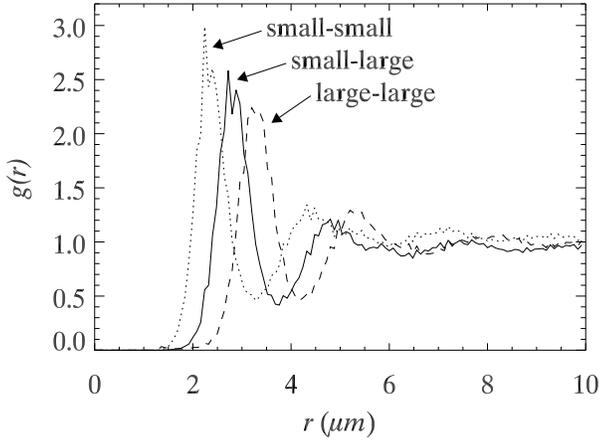}
\end{center}
\caption{The pair correlation function $g(r)$ for a sample with volume
  fraction $\phi=0.57$.  The different curves are for small-small
  pairs, small-large pairs, and large-large pairs, as indicated.
  Due to particle tracking uncertainties ($\pm 0.2$~$\mu$m in
  $x$ and $y$, $\pm 0.3$~$\mu$m in $z$), the measured separation
  $r$ between any pairs of particles has an uncertainty of $\pm
  0.5$~$\mu$m, which significantly broadens the peaks of $g(r)$
  and diminishes their height.
}
\label{fig:rdf_both2001}
\end{figure}

We begin by looking at the structure of the binary sample.  Shown in
Fig.~\ref{fig:rdf_both2001} is the pair correlation function $g(r)$
of a sample with volume fraction $\phi=0.57$.  $g(r)$ relates to
the likelihood of finding a particle a distance $r$ away from a
reference particle.  The three curves correspond to small-small,
small-large, and large-large particle pairs, and their first peak
positions are at approximately $2a_S = 2.36$~$\mu$m, $a_S + a_L =
2.73$~$\mu$m, and $2 a_L = 3.10$~$\mu$m.  The first peaks
of the three curves are fairly broad due to the uncertainty in
locating particle positions, and also in part due to
the particle polydispersity.

\subsection{Dynamical slowing}

\begin{figure}[!ht]
\begin{center}
\includegraphics[width=8.0cm]{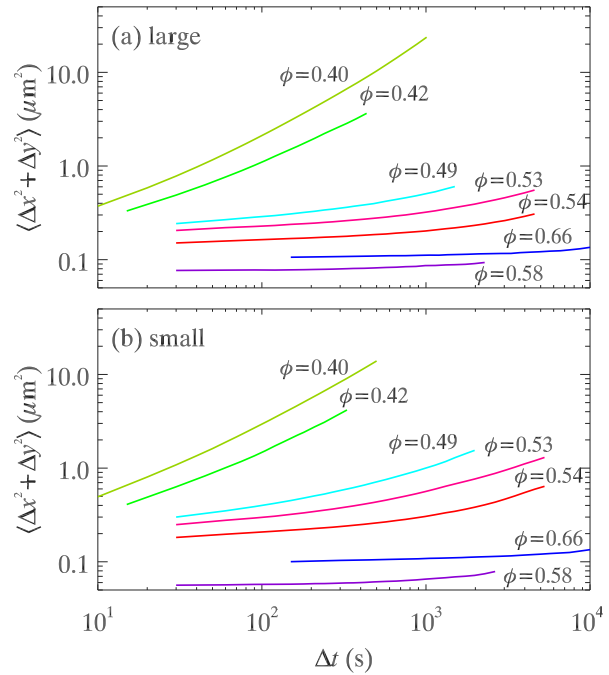}
\end{center}
\caption{(Color online) A log-log plot of mean square displacement
versus time lag for large particles (a) and small particles (b).
Note that our resolution means we cannot accurately measure mean square
displacement values less than 0.1~$\mu$m$^2$, and thus the
plateau height of curves for the highest volume fraction data is
set by this limit, rather than the dynamics.  However, the slight upturn
for those curves at large values of $\Delta t$ is above our
resolution limit and thus real.  Due to
  noise in $\langle \Delta z^2 \rangle$, the data shown are
  $\langle \Delta x^2 + \Delta y^2 \rangle$.
}
\label{fig:msd}
\end{figure}

We wish to show how the motion of particles slows as the volume
fraction increases and approaches the glass transition.  Figure
\ref{fig:msd} shows results of the mean square displacement (MSD)
$\langle \Delta {\vec{r}_{i}}^{2}\rangle$ of large and small particles,
where $\Delta \vec{r}_{i}=\Delta \vec{r}_{i}(\Delta t)$ denotes the
displacement of $i$-th particle in lag time $\Delta t$, and the
brackets indicate an average over all particles and times observed.  Figure
\ref{fig:msd} shows that as the volume fraction increases, particle
motion slows significantly, as expected.  At $\phi=0.4$, small
particles take tens of seconds to move a distance $a_S^2 =
1.4$~$\mu$m$^2$; at $\phi=0.54$ the time has grown to more than
$10^4$~s.  For the lowest volume fraction samples, comparing the two
particle species, we find that $\langle \Delta r^2_S \rangle / \langle
\Delta r^2_L \rangle \approx a_L / a_S$, as expected from the
Stokes-Einstein-Sutherland equation
\cite{einstein1905a,sutherland1905}.

As the volume fraction increases, the MSD plots show a
characteristic ``cage trapping'' plateau.
Particles cannot diffuse freely, but instead are ``caged'' by
their nearest neighbor particles 
\cite{berne97,doliwa98,bartsch98,weeks02,weeks02sub,shattuck07}.
The upturn in the MSD curve is identified with rearrangements
of the cage, allowing the particle to move to a
new location, perhaps caged by different particles.  Although the
smaller particles diffuse faster than the large particles, MSD curves
for both show upturns at similar time scales, indicating that
their dynamics are strongly coupled \cite{lynch08}.
Note that the height of the plateaus (0.1 - 0.2~$\mu$m$^2$) are
larger than in prior work \cite{weeks00}, 
because of the larger particle tracking noise present in this
binary experiment as compared to the prior work with monodisperse
particles.  The noise results in apparent displacements $\langle
\Delta x^2 \rangle = \langle \Delta y^2 \rangle =
(0.2$~$\mu$m)$^2$, independent of $\Delta t$, and so the minimum
reliable value for Fig.~\ref{fig:msd} is limited to
$\sim 0.08$~$\mu$m$^2$.  In particular the difference between the
$\phi=0.58$ and $\phi=0.66$ data is probably not significant, but
due to higher tracking errors for the higher volume fraction
data.  The noise prohibits careful analysis
of the mean square displacement data along the lines of prior
work \cite{ngai98,vanmegen09}.

\begin{figure}[!t]
\begin{center}
\includegraphics[width=8.0cm]{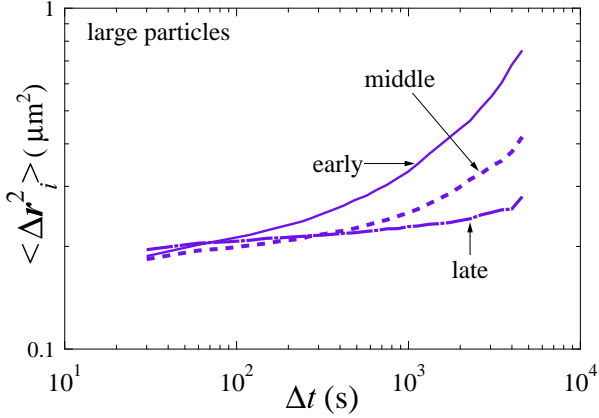}
\end{center}
\caption{(Color online) MSD plot for the large
  particles from a $\phi=0.59$ sample.  The three curves
  correspond to time regimes with 
  $0 < t < 4800$~s, $4800 < t <
  9600$~s, and $9600 < t < 14400$~s.  Note that in the first time
  regime, the sample ages appreciably, so the ``early'' mean square
  displacement data should be interpreted with caution.  The
  time-dependence of $\langle \Delta r^2\rangle$ is clearly seen,
  indicating the presence of aging.  As $\phi=0.59$ is the lowest
  volume fraction in which this behavior is seen, we conclude that the
  glass transition occurs at $\phi_g\approx 0.58$.  Due to
  noise in $\langle \Delta z^2 \rangle$, the data shown are
  $\langle \Delta r^2 \rangle = \langle \Delta
  x^2 + \Delta y^2 \rangle$.
}
\label{fig:age}
\end{figure}

For the samples with $\phi \geq \phi_g \approx 0.58$, the MSD
curves are nearly flat, suggesting that on our experimental time
scales, these samples behave as glasses.  Glasses are fundamentally
non-equilibrium systems, so that physical properties for glasses
depend on the preparation history in general and, in particular,
the time since they were initially formed.  This time-dependence
is known as aging, and can be quantified by examining the
MSD at different times since the start of the experiment
\cite{courtland03,cianci06ssc,lynch08}.  Figure \ref{fig:age}
shows MSD data from $\phi=0.59$.  The trajectory data are broken into three
equal duration segments and the MSD calculated within each segment;
for the older segments, the MSD curve decreases in height (see
the caption for details).  The sample is most active immediately
after being formed, and continues to slow down as time elapses.
The aging of the MSD appears in samples for $\phi\ge 0.59$, while
no samples for $\phi \le 0.58$ show aging.  From the onset of
aging, we conclude that the glass transition point is at volume
fraction $\phi\approx 0.59$, similar to that seen for monodisperse
samples \cite{pusey86}.  Note that our particle size uncertainty of
$\pm 0.02$~$\mu$m (radius) leads to a systematic volume fraction
uncertainty, so our estimate is $\phi_g = 0.59 \pm 0.02$ as a
comparison with other work.

\begin{figure}[!t]
\begin{center}
\includegraphics[width=8.0cm]{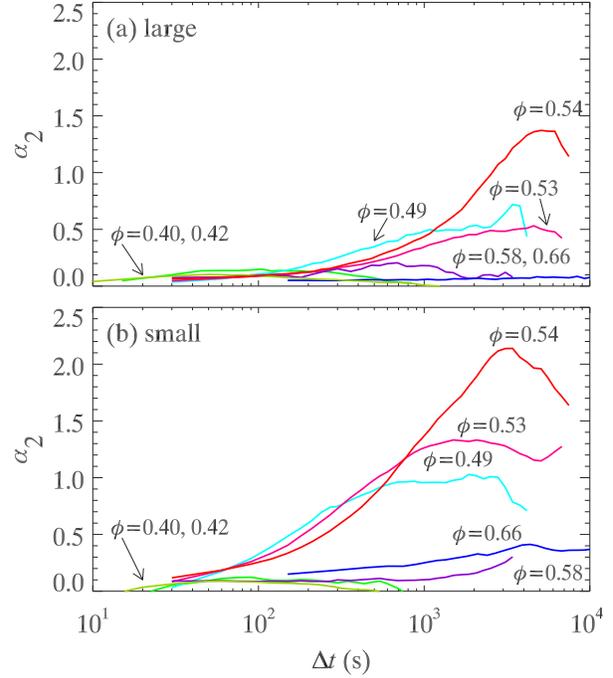}
\end{center}
\caption{(Color online) A semi-log plot of the non-Gaussian parameter
  $\alpha_2$ versus lag time for large particles (a) and small
  particles (b).  The curves for $\phi=0.40, 0.42$ are
  essentially indistinguishable.
}
\label{fig:ngp}
\end{figure}

Particles involved in a cage rearrangement event move significant
distances compared to when they are caged, and prior work noted that
the distribution of displacements is unusually broad on the time scale
of the rearrangement \cite{weeks02,kob97}.  This is quantified by
calculating the non-Gaussian parameter $\alpha_2(\Delta t)$, which for
a one-dimensional distribution of displacements is defined as
\begin{equation}
\alpha_{2}(\Delta t)=\frac{\langle\Delta x^{4}\rangle}{3\langle\Delta
x^{2}\rangle}-1, 
\end{equation}
where $\Delta x=x(t+\Delta t)-x(t)$ denotes the $x$
displacement for time lag $\Delta t$ \cite{rahman64}, and
the angle brackets indicate an average over all particles and all
initial times $t$.  If the
distribution of displacements $\Delta x$ is Gaussian, then $\alpha_2
= 0$ by construction.  If events with large displacements are more
common than would be expected from a Gaussian distribution, then
$\alpha_2 > 0$.  Figure \ref{fig:ngp} shows the results of the
non-Gaussian parameter (NGP) for one-dimension displacements of
large and small particles.  The curves peak at time scales where
cage rearrangements are most important \cite{donati98,weeks00},
and thus coincides with the upturn of the MSD curves.

$\alpha_2$ is fairly sensitive to experimental noise, although
fortunately the ``signal'' this parameter measures comes from
particles moving large distances, which are less susceptible to
particle position uncertainty.  The differences between the data
shown in Fig.~\ref{fig:ngp} and the data of Ref.~\cite{weeks00} are
probably more due to noise and uncertainties in measuring $\phi$,
rather than systematic differences between our binary experiments
and the prior monodisperse experiments \cite{weeks00}.  The peak
heights of $\alpha_2$ for our small particles are similar to those
seen in Ref.~\cite{weeks00}, for roughly similar volume fractions.
One notable difference is that Ref.~\cite{weeks00} found that for
glassy samples, $\alpha_2$ started high (1 - 2) and decreased
steadily, whereas in our current data, $\alpha_2$ remains fairly
small for all $\Delta t$, with unclear dependence on $\Delta
t$.  This is probably due to our noise, as for glassy samples, 
few particles move distances large enough to be outside of our
particle tracking uncertainty.

Figure~\ref{fig:ngp} also reveals that the motions of the small
particles are more dynamically heterogeneous, with the maximum NGP peaking
above 1.5 for the small species but only reaching 0.8 for the
large species.  This is consistent with recent observations
of aging binary colloidal glasses, which likewise found the small
particles had more non-Gaussian motion \cite{lynch08}.

From Figs.~\ref{fig:msd} and \ref{fig:ngp} we conclude that the
dynamics of the large and small particles are qualitatively the same,
although with small quantitative differences.  In particular, the
time scale over which particles escape cages is the same for both, as
is the time of peak non-Gaussianity.  In much of the subsequent
analysis, therefore, we consider both species together in order to
obtain better statistical validity.

\subsection{Local environment influences mobility}

We wish to understand the origins of dynamical heterogeneity.
For a hard-sphere system, or an overdamped system such as
our experimental colloidal suspension, the only variable is
the local structure.  Clearly structure has some relation with
particle mobility \cite{harrowell04,widmercooper05}, although this
relationship may be difficult to see and not directly predictive
in nature \cite{conrad05}.  Prior work found that more disordered
environments are weakly correlated with higher particle mobility
\cite{weeks02,cianci06ssc}, and a recent study of aging binary
colloidal glasses found a relation between the local composition
and the mobility \cite{lynch08}.

\begin{figure}[!ht]
\begin{center}
\includegraphics[width=6.0cm]{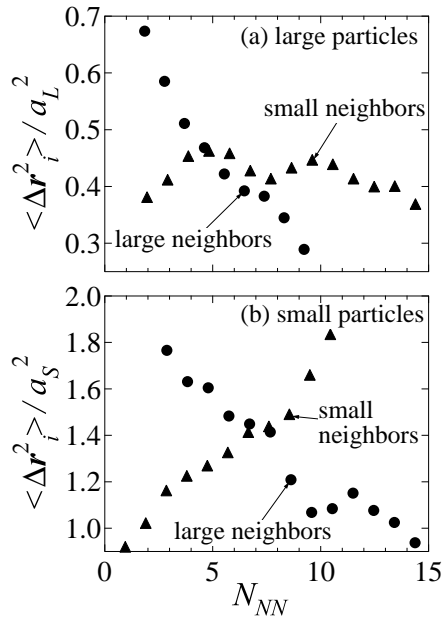}
\end{center}
\caption{Large particle mobility as a function of the number of large
  and small nearest neighbors $N_{NN}$. The panel (b) shows small particle
  mobility. Mobility is very sensitive to
  the number of large neighbors, decreasing sharply as the number of
  large neighbors increases. The number of smaller neighbors has a
  weaker, yet measurable, impact.  These data are for volume fraction
  $\phi=0.53$, using a time scale $\Delta t=3780$~s to define
  displacements.  }
\label{fig:neighbors}
\end{figure}

We quantify a particle's local environment by counting its nearest
neighbors $N_{NN}$, defined as particles closer than the first minimum
of the pair correlation function for the large particles, 4.1~$\mu$m
(Fig.~\ref{fig:rdf_both2001}), and distinguish between large and small
neighbors.  Figure \ref{fig:neighbors} shows that the number of
neighbors of a given type has a strong influence on the mobility of a
particle.  Particles with more large neighbors have, on average, a
lower mobility, while those with more small neighbors a larger
mobility.  These observations agree with studies of aging in binary
colloidal glasses, \cite{lynch08} and are reminiscent of prior
rheological observations of binary suspensions
\cite{hoffman92,mewis94,vanmegen01} which noted that binary mixtures
have lower viscosities than single-component samples with equivalent
total volume fraction.  The reasoning is that binary suspensions can
in general be packed to higher volume fractions, and so have more free
volume than monodisperse samples at the same volume fraction.  Figure
\ref{fig:neighbors} suggests that the small particles indeed
``lubricate'' large particles, as previously proposed
\cite{vanmegen01}. Conversely, large neighbors significantly inhibit
the motion of both large and small particles.  The lubricant effect
for large particles (which have less free volume) is less pronounced,
agreeing with prior observations of monodisperse suspensions which
found that regions of mobility correlated with regions of larger free
volume \cite{weeks02}.

\subsection{Cooperative motions}\label{sub:CooperativeMotion}

Prior work has shown that the higher mobility molecules in a
supercooled liquid are 
distributed in a
spatially heterogeneous fashion \cite{ediger00,donati98}.  In
monodisperse colloidal systems, direct imaging using microscopy found
that particles rearrange in cooperative groups
\cite{marcus99,weeks00,kegel00}.  Following the prior work, we
characterize the cooperative nature of colloidal rearrangements by
studying the dynamics over a time scale $\Delta t^*$ that corresponds
to the maximum of the NGP \cite{donati98,weeks00}.  The maximum
displacement of a particle over that time $D_i$ is defined as
\begin{equation}
D_i(t):=\max_{t,t+\Delta
t^{*}} (|\vec{r}_{i}(t_{2})-\vec{r}_{i}(t_{1})|)
\end{equation}
where $\max_{t,t+\Delta t^{*}}(X)$ is the maximum value of $X$ using
times $t_1$, $t_2$ such that $t\le t_{1} < t_{2}\le t+\Delta t^{*}$.
Taking the maximum displacement results in a quantity that is less
sensitive to random Brownian motion than the ordinary displacement
$\Delta r$.  Following prior work, \cite{donati98,weeks00} a threshold
$D^{*}(\phi)$ is chosen such that on average, 5\% of the particles at
any given time have $D_i(t) > D^*$.  These particles are termed
``mobile particles'' and generally are the ones undergoing cage
rearrangements.  (Note that at any particular time, the fraction of
particles matching this definition is not required to be 5\%
\cite{dinsmore01}.)

\begin{figure}[!t]
\begin{center}
\includegraphics[width=6.0cm]{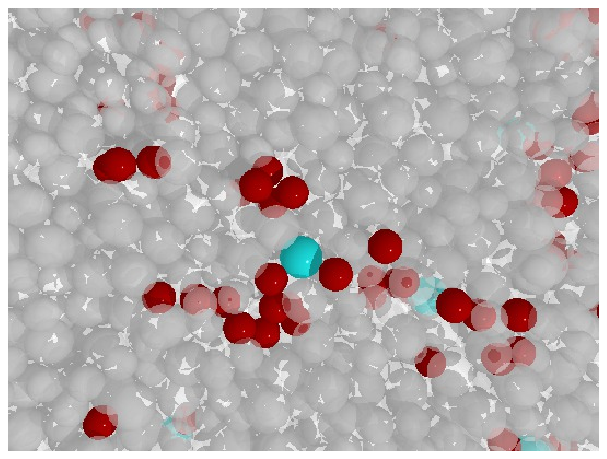}
\end{center}
\begin{center}
\includegraphics[width=6.0cm]{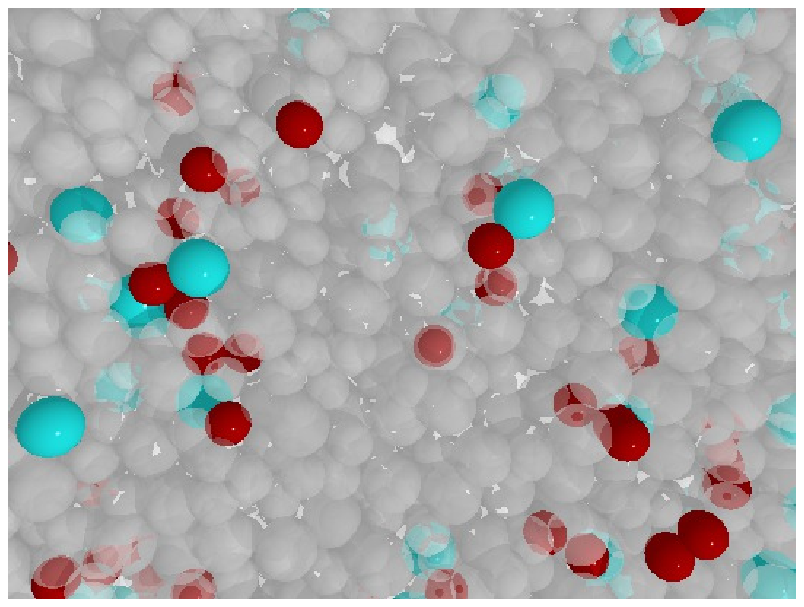}
\end{center}
\caption{(Color online) Snapshots of system for $\phi=0.54$ (upper) and
  $\phi=0.66$ (bottom). The red (dark gray) spheres are small
mobile particles, and the blue (light gray) spheres are large
mobile particles.
  Mobile particles are defined as those making the largest displacements
  at this particular moment in time;
  see text for further details.  We set the time lag
  for the displacement as the cage breaking time scale (the peak time
  of the NGP $t_{\text{NGP}}$) which is $\Delta t^{*}=3000$~s for 
  $\phi=0.54$ and  $t^{*}=4000$~s for $\phi=0.66$.
}
\label{fig:cluster_picture}
\end{figure}

Figure \ref{fig:cluster_picture} shows snapshots of our system,
highlighting the mobile particles.  Clusters of these mobile
particles are visible, in agreement with previous work which
found similar mobile regions \cite{donati98,weeks00,lynch08}.
The clusters are somewhat smaller than those seen previously in
single-component colloidal suspension \cite{weeks00}; apparently
the dynamics in binary mixtures are less spatially heterogeneous.
Our result is in agreement with the results of a simulation
study for polydisperse hard-disk systems \cite{kawasaki07},
which found that polydispersity reduces dynamic heterogeneity.
It is also apparent in Fig.~\ref{fig:cluster_picture} that the
small particles dominate the motions for the lower volume fraction
(top, $\phi=0.54$) whereas the two species contribute more equally
in the glassy sample (bottom, $\phi=0.66$).

\subsection{Length scales of spatial dynamical heterogeneity}

Pictures such as Fig.~\ref{fig:cluster_picture} are qualitative
evidence of dynamical heterogeneity.  For quantitative information, we
consider the vector and scalar spatial-temporal correlation functions
\cite{doliwa00} $S_{vec}(R,\Delta t)$ and $S_{scl}(R,\Delta t)$ defined as
\begin{eqnarray} 
S_{vec}(R, \Delta t) & := & 
   \frac{\left<\Delta\vec{r}_{i}\cdot
\Delta\vec{r}_{j}\right>_{\text{pair}}}{\left<\Delta \vec{r}^{2}\right>} 
\label{d_correlation} \\
S_{scl}(R, \Delta t) & := & 
    \frac{\left<\delta r_{i} \delta r_{j}\right>_{\text{pair}}}{\left<(\delta r)^{2}\right>}.
\label{m_correlation}
\end{eqnarray}
The vector function $S_{vec}(R,\Delta t)$ characterizes
correlations in the vector displacements $\Delta \vec{r}_i =
\vec{r}_i(t+\Delta t)-\vec{r}_i(t)$; the similar function
$S_{scl}(R,\Delta t)$ uses the scalar displacement $\delta
r_{i}=|\Delta\vec{r}_{i}|-\left<|\Delta\vec{r}_{i}|\right>$.
The angle brackets $\langle \rangle$ denote an average over
all particle pairs with separation $R$ at initial time $t$
as well as an average over $t$.  The denominators of both
correlation functions are averaged over all particles and time,
and do not depend on $R$. The correlation function defined by
eqn~(\ref{d_correlation}) indicates a vector correlation, and that
defined by eqn~(\ref{m_correlation}) a scalar correlation. If
particles correlate perfectly, the correlation functions
are unity.  These correlation functions give information about
spatial correlations for fixed lag time $\Delta t$, and about
temporal dependence of the correlations for fixed separation $R$.
We calculate these functions for all pairs of particles, without
concern for the particle sizes, both to improve our statistics and
because we do not find significant differences for large and small
particles only.  Most of the ``signal'' of correlated motion comes
from the particles undergoing larger than average displacements,
and so the results are less sensitive to the particle tracking
uncertainties than the mean square displacement.


\begin{figure}[!t]
\begin{center}
\includegraphics[width=8.0cm]{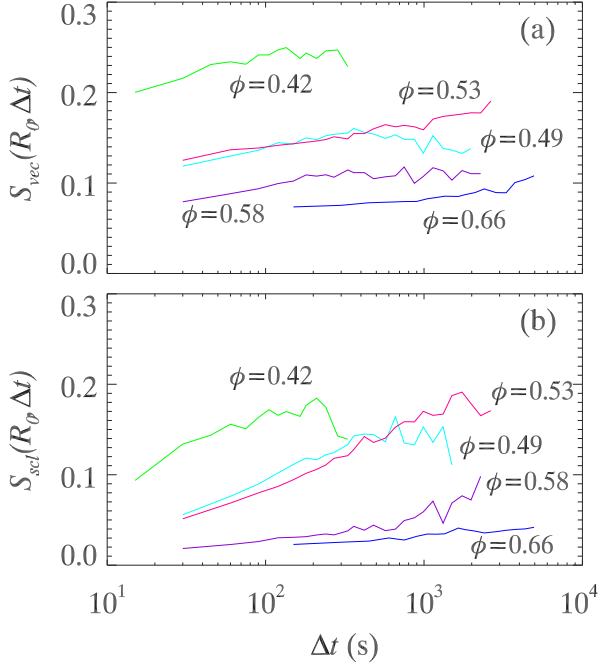}
\end{center}
\caption{(Color online) Plot of the correlation functions in
  which the distance $R_0=a_S+a_L=2.73$~$\mu$m is set at the sum of
  the particle radii.
  (a) represents
  the vector correlation [eqn~(\ref{d_correlation})], and (b) the
  scalar correlation [eqn~(\ref{m_correlation})]. 
}
\label{fig:cor_time_both}
\end{figure}

Figure \ref{fig:cor_time_both} shows the lag time dependence of
these correlation functions, in which the distance $R$ is set as the
first peak distance of the small-large pair correlation function
$g(r)$ (the solid line in Fig.~\ref{fig:rdf_both2001}).  At time
scales larger than those shown in Fig.~\ref{fig:cor_time_both}, the
results become too uncertain, due to lack of data.  In intermediate
volume fraction region ($\phi < \phi_g$), both correlation functions
increase with $\Delta t$.  For the two lowest volume fractions
($\phi=0.42, 0.49$), the correlation functions eventually decrease
at large $\Delta t$, but our data do not extend to large enough
$\Delta t$ to see this for higher volume fractions.  Overall, in
conjunction with Fig.~\ref{fig:msd}, Fig.~\ref{fig:cor_time_both}
suggests that larger motions are more correlated with the motions
of their neighboring particles.  This agrees with prior experiments
\cite{weeks00,weeks07cor}.  The amplitude of the
correlation decreases as $\phi$ increases, with the exception of
the $\phi=0.49, 0.53$ data which are similar.

For glassy samples ($\phi > 0.6$) the correlation functions 
are small, suggesting that there is little correlation of the
motion of neighbors.  This is both because there is little overall
motion in glassy samples (see Fig.~\ref{fig:msd}) and also the
motion that does occur is dominated by Brownian motion within the
cage, which is less correlated that the motions responsible for
cage rearrangements \cite{weeks02}.  Furthermore, it is probably
erroneous to even consider time-averaged correlation functions for
glassy samples, as in this current work the dynamics slow with
time (Fig.~\ref{fig:age}), and so a time average is of dubious
validity.   (Earlier work studied well-aged samples where the
dynamics were only slow aging, and thus a time average was more
sensible \cite{weeks07cor}.)

\begin{figure}[!ht]
\includegraphics[width=0.92\linewidth]{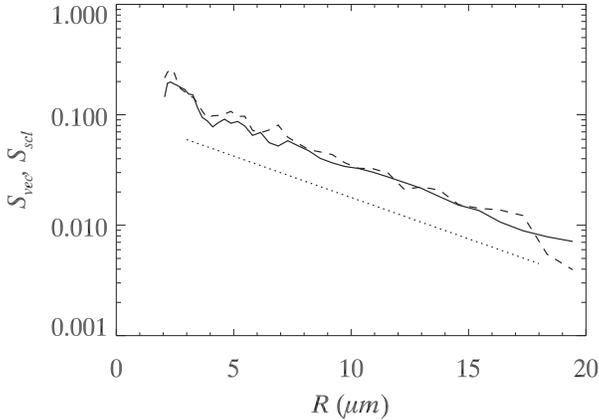}
\caption{
Semi-log plot of the spatial correlation functions of
$\phi=0.53$, where the time lag is set as $\Delta t_{\text{NGP}}=2000$s. 
The solid line is $S_{vec}(R,t_{\text{NGP}})$ and the
dashed line is $S_{scl}(R,t_{\text{NGP}})$.
The dotted line represents an exponential function with
a decay length of 5.8~$\mu$m, a good fit to both functions in
this particular case.
}
\label{fig:cordr}
\end{figure}

To consider the spatial dynamical heterogeneities, we
plot the correlation functions as a function of $R$ in
Fig.~\ref{fig:cordr} (for $\phi=0.54$; results for other
volume fractions are similar).  For small separations around
$R=3.5$~$\mu$m, there is a dip in the correlation functions, which
corresponds to the dip in the small-large pair correlation function
at the same position (solid line in Fig.~\ref{fig:rdf_both2001});
the peak around $R=2.8$~$\mu$m likewise corresponds to the peak
of the small-large pair correlation function.  Thus, a particle's
motion is correlated with that of its nearest neighbors, while
particles separated by a less structurally favorable distance are
less likely to have correlated motion.

\begin{figure}[t]
\begin{center}
\includegraphics[width=8.0cm]{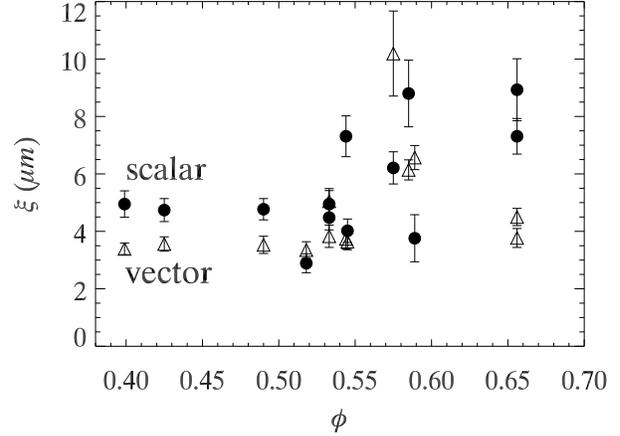}
\end{center}
\caption{(Color online) The relationship between the length scales and
  the volume fraction.  Shown are the length scales for the
vector correlation function (open triangles) and the scalar
correlation function (closed circles).
The symbols indicate the average value, and the
  error bars show the range of values found for different lag times
  $\Delta t$.  These length scales are extracted from the 
  correlation function for all particles (large + small).
  }
\label{fig:sl_both}
\end{figure}

We fit our data with an exponential function $S\simeq A\exp(-R/\xi)$
and extract the decay length $\xi$.  Figure \ref{fig:sl_both}
shows both the vector (triangles) and scalar (circles) decay
lengths as a function of the volume fraction.  The length scales
are essentially constant until close to $\phi_g \approx 0.58$,
when they show a sharp increase.  Our data are too noisy to
draw conclusions about how the length scales grow near $\phi_g$,
although simulations of binary Lennard-Jones mixtures did not find
a divergence \cite{narumi08}.  The largest length scale seen is
$\approx 10$~$\mu$m~$\approx 8a_S \approx 6.5 a_L$, similar to
prior studies of monodisperse colloids \cite{weeks07cor}.  For
$\phi > \phi_g$, the scalar length seems large and the vector
length decreases, although as noted above, the data should be
treated with caution as the samples are aging.

\subsection{Temporal dynamical heterogeneity}

The prior subsection showed that the motion of colloidal particles
in our dense samples are spatially heterogeneous.  We now study
their temporal heterogeneity, using the four-point susceptibility
$\chi_4$ which measures the correlation in dynamics between any two
points in space within some time window~\cite{glotzer00_2,keys07}.
The actual value of $\chi_4$ is a measure of the average number
of particles whose dynamics are correlated, which in turn relates
back to the spatial heterogeneity~\cite{brambilla09}.

\begin{figure}
\includegraphics[width=3.2in]{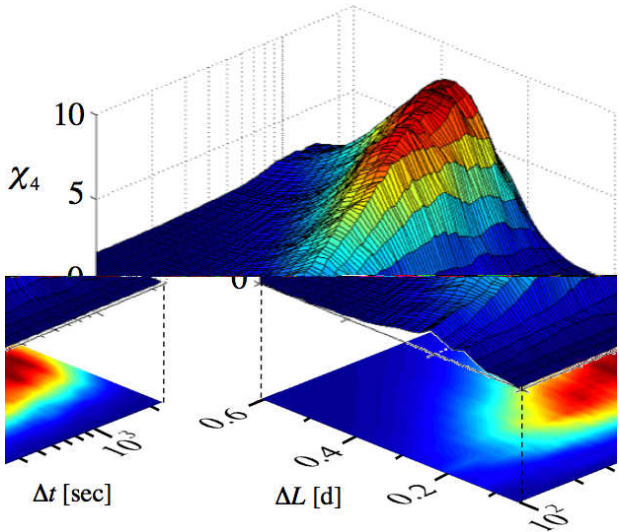}
\caption{(Color online) Surface plot of $\chi_4$ for large particles
within a sample with $\phi = 0.52$.}
\label{fig:chi4}
\end{figure}

Here we only compute the self contribution to $\chi_4$, since it has
been shown to be the dominating term~\cite{glotzer00_2,keys07}. The
self part is computed from temporal fluctuations in particle mobility,
where a particle is defined to be mobile if its displacement over some
time interval $\Delta{}t$ is larger than some threshold distance
$\Delta{}L$~\cite{keys07}. Using this definition, each particle at
each time can be labeled mobile or immobile and the fraction of mobile
particles $Q(t)$ can be computed for each frame recorded. $Q(t)$
varies from frame to frame due to the spatial heterogeneity. The
temporal fluctuations in $Q(t)$ are quantified by the self part to
$\chi_4$ and written as 
\begin{equation}
    \chi_4 = N[\langle Q(t)^2
    \rangle_t - \langle Q(t) \rangle_t^2 ],
\label{eq:chi4}
\end{equation}
where $N$ is the number of particles. $N$ also varies from frame
to frame as particles move in and out of the field of view;
we average $N$ over all frames and use $\langle N \rangle$ in
eqn~(\ref{eq:chi4}). (The factor of $N$ arises because the variance
scales inversely with particle number.)  Note that $\chi_4$
measures temporal fluctuations in mobility without regard for
the spatial correlations between mobile particles, whereas
the correlation functions $S_{vec}$ and $S_{scl}$ studied in
the previous section measured spatial correlations of mobility
without regard for the temporal correlations.  In a sense, then,
these two methods of analysis are complementary.

From eqn~(\ref{eq:chi4}) it's evident that $\chi_4$ will depend on our
choice of $\Delta{}L$ and $\Delta{}t$ as shown in Fig.
\ref{fig:chi4}, where $\chi_4$ is plotted for the larger particles
within a $\phi = 0.52$ sample for various values of $\Delta{}t$ and
$\Delta{}L$. This plot shows that $\chi_4$ is characterized by a
function that has a maximum at ($\Delta{}t_{\max}$,
$\Delta{}L_{\max}$). This maximum in $\chi_4$ indicates a typical
timescale $\Delta{}t_{\max}$ where the dynamics are most
heterogeneous, and likewise $\Delta L_{\max}$ indicates a typical
length scale distinguishing caged motions from cage rearrangements.

\begin{figure}
\includegraphics[width=3.3in]{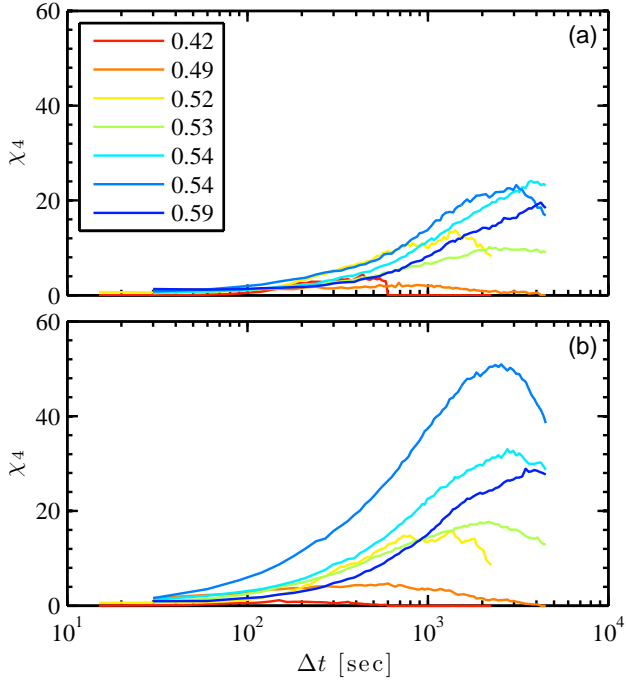}
\caption{(Color online) (a) is a plot of the susceptibility of large
particles for various packing fractions, and (b) is a plot of
the susceptibility of small particles for various packing fractions.}
\label{fig:chi4_vs_phi}
\end{figure}

Figure \ref{fig:chi4_vs_phi} shows plots of $\chi_4(\Delta{}t,
\Delta{}L=\Delta{}L_{\max})$ for the larger (a) and smaller (b)
particles.  The value of $\chi_4$ is larger in magnitude for the
smaller particles regardless of $\phi$, demonstrating that the
dynamics of the smaller particles are more temporally heterogeneous.
Prior work by Lynch \textit{et al.}~\cite{lynch08} showed a similar
relative mobility; our results build upon this by showing that
smaller particles also experience larger fluctuations, and thus
exhibit more anomalous spatial and temporal behavior.  We also
see that $\chi_4$ grows in amplitude as $\phi$ increases, but
then drops for the glassy sample ($\phi=0.59$), matching the
results of the prior subsections where changes were seen at the
glass transition.

The plots in Fig.~\ref{fig:chi4_vs_phi} all show a maximum in $\chi_4$
at a well defined $\Delta{}t_{\max}$, and that $\Delta{}t_{\max}$ for
the various volume fractions occur at timescales close to where
$\alpha_2$ shows a maximum in Fig.~\ref{fig:ngp} and caging
rearrangements become prominent (the ``knee'' in
Fig.~\ref{fig:msd}). The coincidence of maxima in $\chi_4$ and
$\alpha_2$ suggests that local cage rearrangements are the largest
contributor to the temporal fluctuations. Since small particles show
larger fluctuations, we infer that they may be largely responsible for
facilitating local rearrangements, in agreement with the findings of
Lynch \textit{et al.}~\cite{lynch08}. 

\begin{figure}[t]
\begin{center}
\includegraphics[width=7.5cm]{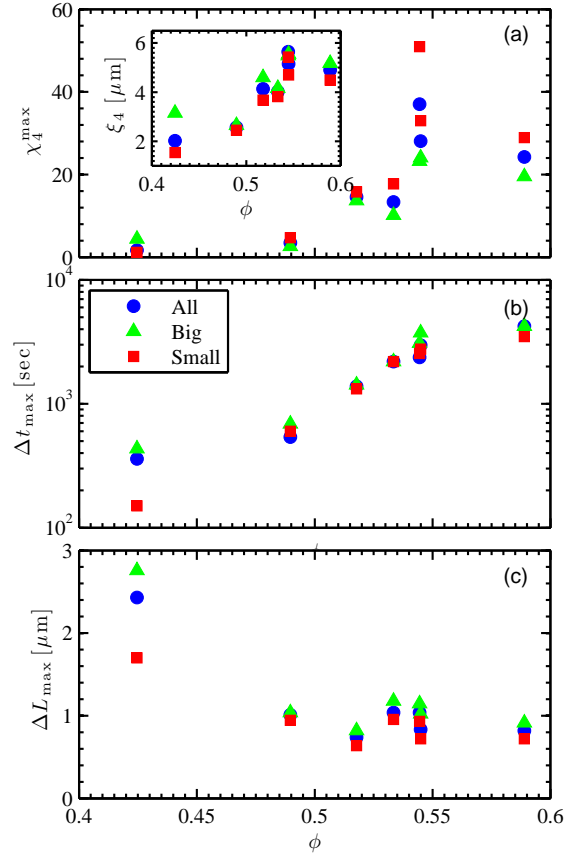}
\end{center}
\caption{
(Color online)
(a) Plot of the maximum of $\chi_4$ as a function of $\phi$,
showing how temporal heterogeneity increases as $\phi \rightarrow
\phi_g \approx 0.58$.
The inset in (a) shows the dependence of the dynamic
heterogeneity length scale $\xi_4 = (\chi_4^{\rm max})^{1/3}$ on $\phi$.  
(b) Plot of the dynamic heterogeneity time scale as a function of
$\phi$.
(c) Plot of the length scale $\Delta L$ as a function
of $\phi$.  For all panels, the symbols are as indicated in
the legend of panel (b).
}
\label{fig:chi4tau}
\end{figure}

$\chi_4^{\max}$, $\Delta{}t_{\max}$, and $\Delta
L_{\max}$ all vary with $\phi$; this dependence is shown in
Fig.~\ref{fig:chi4tau}. Both $\chi_4^{\max}$ and $\Delta{}t_{\max}$
show an increase with $\phi$ illustrating that upon approaching
the glass transition the dynamic heterogeneity and the associated
time scale increases. The increasing time scale also suggests that
local rearrangements take longer at higher $\phi$, in agreement
with the extended plateau at higher $\phi$ in Fig.~\ref{fig:msd}.
The characteristic length scale $\Delta L$ decreases.  This is
in excellent agreement with prior work, which showed that
the displacements for cage rearrangements are smaller as the
glass transition is approached \cite{weeks02}.  In other words,
it requires a smaller displacement to be an anomalously mobile
particle.  This can also be seen by comparing Figs.~\ref{fig:msd}
and \ref{fig:ngp}:  for samples with larger $\phi$, the mean
square displacement has a smaller value even when the non-Gaussian
parameter is large, showing that the distribution of displacements
is overall narrower despite the relatively large fraction of
larger-than-expected displacements.

Using the $\chi_4^{\max}$ data in Fig.~\ref{fig:chi4tau}(a) a
correlation length scale can be estimated by assuming that the
correlations $\chi_4$ measures are correlated particles forming
compact clusters. Since $\chi_4$ is the average number of correlated
particles, then $\chi_4^{\max} = (4/3)\pi\xi_4^3$, where $\xi_4$ is
the radius of the cluster of correlated particles in units of particle
diameters $d$~\cite{keys07, sarangapani08b}. The inset in
Fig.~\ref{fig:chi4tau}(a) shows the dependence of $\xi_4$ on
$\phi$. Similarly as with the relaxation time, we see a tendency in
$\xi_4$ to increase with $\phi$. The growth in $\xi_4$ is about a
factor of 4 when the volume fraction is increased from a liquid to a
dense supercooled state. Our values of $\xi_4$ are roughly the same as
those measured in a 2D fluidized granular bed on approaching the
jamming point~\cite{keys07}. When compared to $\xi$ shown in
Fig.~\ref{fig:sl_both} the diameter of these correlated clusters
2$\xi_4$ is roughly the same size.

The time scales $\Delta{}t_{\max}$ are
analogous to the $\alpha$ relaxation time scales measured in
molecular supercooled liquids. In many cases the $\alpha$ relaxation
time scales are well described using either a Vogel-Fulcher-Tammann
(VFT) model or Mode-Coupling Theory (MCT). 

The first model, VFT, predicts that the time scales should
obey the form 
\begin{equation}
\label{vfteqn}
\Delta{}t_{\max} = \Delta{}t_0\exp(E/(1 - \phi/\phi_0)),
\end{equation}
 where $\Delta{}t_0$, $E$, and $\phi_0$ are all fitting parameters. In
 the model $\Delta{}t_0$ is an attempt time to undergo relaxation
 events over some typical length scale.  For our experiment, this
 length scale would be on the order of a particle diameter and the
 attempt time would be the time it takes a particle to diffuse over
 this length scale in the dilute limit.  Using the
 Stokes-Einstein-Sutherland formula and a viscosity of $2.18$
 mPa$\cdot$s (measured for the fluid in absence of colloids) we
 estimate that at room temperature it should take the small particles
 about 11 seconds and the large particles 25 seconds to diffuse their
 own diameter \cite{einstein1905a,sutherland1905}.  The fitting
 parameter $\phi_0$ is the packing fraction at which diffusive motion
 should cease. This should occur at random close packing of $\phi \sim
 0.65$ (using the value appropriate for our binary suspension).
 However, as pointed out by Brambilla
 \textit{et.~al}~\cite{brambilla09}, there is a debate as to whether
 the divergence predicted by VFT should occur at the jamming point or
 at a slightly different packing fraction. To definitively show if
 this is the case one would need very careful measurements extremely
 close to the jamming point which is beyond the scope of this
 paper. The final fitting parameter is $E$, the fragility, which is a
 material dependent value.  The fragility is a measure of how
 sensitive the time scale is to small changes in volume fraction.  For
 a molecular system $E$ measures how sensitive the relaxation time is
 to small changes in temperature, and $E$ ranges between $\approx$
 1-100.  Materials with low $E$ values are termed fragile glass
 formers and those with large $E$ values termed strong glass
 formers~\cite{angell95}.

\begin{figure}
\includegraphics[width=3.4in]{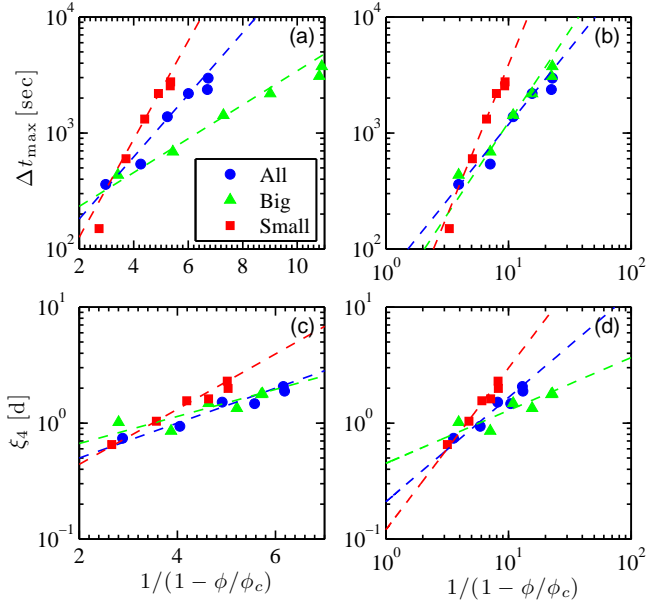}
\caption{
(Color online)
(a) Log-linear and (b) log-log plots of temporal dynamical
heterogeneity time scale, with fits to eqn~(\ref{vfteqn}) in panel
(a) and eqn~(\ref{mcteqn}) in panel (b).  (c) Log-linear and
(d) log-log plots of the $\chi_4$ length scales, with fits to
eqn~(\ref{vfteqn}) in panel (c) and eqn~(\ref{mcteqn}) in panel (d).
}
\label{fig:scalinglaws}
\end{figure}

The second model, MCT, predicts a scaling of 
\begin{equation}
\label{mcteqn}
\Delta{}t_{\max} = \Delta{}t_0(1 - \phi/\phi_c)^{\gamma},
\end{equation}
where $\Delta{}t_0$, $\gamma$, and $\phi_c$ are the fitting
parameters~\cite{Gotze99}. $\phi_c$ in this model takes
a different meaning with the divergence predicted to occur
near the glass transition volume fraction, not at random close
packing. In light scattering experiments performed by Brambilla
\textit{et. al} on 10\% polydisperse colloidal samples they found
$\phi_c \approx 0.59$, slightly above the glass transition volume
fraction~\cite{brambilla09}. Their work also showed that near the
divergence point the dynamics deviate from the predicted form, but
that in the supercooled regime the MCT equation describes the data
well. They also found a scaling exponent of $\gamma = 2.5 \pm 0.1$.

\begin{table}
\small
\caption{This table displays the fitting parameters found when
fitting the data to either a VFT scaling or a power law scaling. The
uncertainties of the fitting parameters are found by adjusting the fitting
parameters until they no longer provide reasonable fits.}
\label{table:fittingvalues}
\begin{center}
\begin{tabular}{cccc}
\hline
VFT: & $\Delta{}t_0$ [sec] or $\xi_4^0$ [d] & $E$ & $\phi_0$ \\
\hline
All $\Delta{}t_{\max}$ & $70 \pm 50$ & $0.6 \pm 0.3$ & $0.64 \pm 0.03$ \\
Big $\Delta{}t_{\max}$ & $200 \pm 160$ & $0.4 \pm 0.2$ & $0.6 \pm 0.03$ \\
Small $\Delta{}t_{\max}$ & $25 \pm 20$ & $0.4 \pm 0.35$ & $0.67 \pm 0.03$ \\
All $\xi_{4}$ & $0.3 \pm 0.2$ & $0.4 \pm 0.2$ & $0.65 \pm 0.05$ \\
Big $\xi_{4}$ & $0.5 \pm 0.2$ & $0.3 \pm 0.2$ & $0.66 \pm 0.07$ \\
Small $\xi_{4}$ & $0.2 \pm 0.15$ & $1.0 \pm 0.8$ & $0.68 \pm 0.07$ \\
\hline
\multicolumn{4}{c}{ } \\
\hline
Power law: & $\Delta{}t_0$ [sec] or $\xi_4^0$ [d] & $\gamma$ or $\delta$ & $\phi_c$ \\
\hline
All $\Delta{}t_{\max}$ & $100 \pm 90$ & $1.3 \pm 0.6$ & $0.57 \pm 0.02$ \\
Big $\Delta{}t_{\max}$ & $90 \pm 70$ & $1.6 \pm 0.8$ & $0.57 \pm 0.02$ \\
Small $\Delta{}t_{\max}$ & $20 \pm 10$ & $2.6 \pm 0.9$ & $0.61 \pm 0.01$ \\
All $\xi_{4}$ & $0.2 \pm 0.1$ & $0.9 \pm 0.4$ & $0.59 \pm 0.03$ \\
Big $\xi_{4}$ & $0.5 \pm 0.1$ & $1.4 \pm 0.5$ & $0.62 \pm 0.04$ \\
Small $\xi_{4}$ & $0.15 \pm 0.05$ & $0.4 \pm 0.2$ & $0.57 \pm 0.02$ \\
\hline
\end{tabular}
\end{center}
\end{table}%

Fits to the measured time scales using the two fitting
models are shown in Fig.~\ref{fig:scalinglaws}(a) and
(b), and the corresponding fitting parameters are shown in
Table~\ref{table:fittingvalues}. In the previous paragraphs
reasonable fitting values were given for some of the different
fitting parameters.  The $\Delta t_0$ values are significantly
larger than the dilute concentration diffusive time scales,
for both the VFT and MCT fits, although the agreement is off
by only a factor of two for the small particles (25~s for VFT,
20~s for MCT, and $\tau_D = 11$~s).  For the VFT fit, $\phi_c$ is
near $\phi_{rcp}$ as predicted. For the MCT fit $\phi_c$ is near
the expected glass transition volume fraction of $\approx$ 0.58.
The MCT exponent $\gamma$ is smaller than that found by Brambilla
\textit{et.~al}~\cite{brambilla09}, who found $\gamma=2.5$, with
the exception of the small particles for which we find $\gamma=2.6
\pm 0.9$.

Our data gives
fragilities on the order of 0.5, consistent with fragility values
from a study of a 2D fluidized granular bed~\cite{keys07}. When
compared to a molecular system our colloidal system would be
considered a very fragile glass former.

In the study on the motion of grains in a 2D fluidized granular
bed it was shown that the length scales can also be fitted well
to the models used to fit the time scales where the VFT formula
becomes $\xi_{4} = \xi_4^0\exp(E/(1 - \phi/\phi_c))$ and the MCT
formula becomes $\xi_{4} = \xi_4^0(1 - \phi/\phi_c)^{\delta}$,
where $\delta$ in work by Berthier \textit{et.~al} is
predicted to be 2/3~\cite{berthier07a, berthier07b}. The work of
Brambilla \textit{et. al} found that $\delta = 2/3$ fitted their
light scattering data very well~\cite{brambilla09}.

The fits to the length scales are shown in
Fig.~\ref{fig:scalinglaws}(c) and (d), and the fitting values are
shown in Table~\ref{table:fittingvalues}. The fitting values found
for the VFT fits are physically feasible where the fragilities and
divergence points compare well to the fitting parameters previously
found for the VFT fits to the time scales. The MCT fits are also
reasonable, although our scaling exponents $\delta$ is only
consistent with the predicted value of 2/3 due to our large error
bars.  The MCT divergence at $\phi_c$ is close to $\phi_g$, as
expected.

With the range of volume fractions presented in this paper we can not
conclusively show which model fits better. Both models capture and
predict the time and length scales associated with dynamic
heterogeneity, and the derived fitting parameters of both compare well
to expected values.

\section{Summary} \label{sec:summary}

We have used confocal microscopy to study three-dimensional motion of
particles in binary colloidal mixture.  The volume fraction $\phi$ is
varied from 0.4-0.7 and a glass transition, characterized by aging
dynamics, is found at $\phi\approx 0.58$.  The dynamics of large and
small particles are qualitatively similar.  At volume fractions
approaching the glass transition, both show an increase in motion at
the same characteristic cage breaking time scale.  This time scale
also corresponds with the time over which the displacement
distribution functions are broadest (most non-Gaussian).
Particle motion is
facilitated by the presence of small neighbors, and inhibited by large
neighbors, consistent with the idea that small particles serve as
lubricants.  We have investigated vector and scalar correlation
functions and extracted specific length scales associated with the
spatial decay in correlation of the displacements.  This length
slightly increases with volume fraction, although it does
not appear to diverge as the glass transition is approached.
The temporal correlations also give rise to length scales and
time scales which grow as the glass transition is approached,
although the form of this growth is ambiguous with respect to
power-law or exponential growth.

The presence of particle tracking noise makes certain
measurements more difficult, in particular, the pair correlation
functions (Fig.~\ref{fig:rdf_both2001}) and the mean square
displacement (Fig.~\ref{fig:msd}).  The primary conclusions
of our work, however, focus on the particles that move large
distances, and these measurements have a ``signal'' (the
distance moved) larger than the ``noise'' (the instantaneous
positional uncertainty).  Our measurements of the non-Gaussian
parameter (Fig.~\ref{fig:ngp}, spatial correlation functions
(Figs.~\ref{fig:cor_time_both},\ref{fig:cordr}), correlation lengths
(Fig.~\ref{fig:sl_both}, and dynamic susceptibility measurements
(Figs.~\ref{fig:chi4} - \ref{fig:scalinglaws}) are robust to
the noise.  Likewise, the identification of nearest neighbors is
fairly robust to even moderate fluctuations in pair-wise particle
separations, and so we have confidence in our data showing that
having fewer large neighbors enhances a particle's mobility
(Fig.~\ref{fig:neighbors}).

\section{Acknowledgments}
We thank the Soft Condensed Matter Group of the Physics
Department in Emory University for discussions, in
particular J.~M.~Lynch and G.~C.~Cianci.  T.~N.~was supported by the
21st Century COE Program ``International COE of Flow Dynamics''
of Tohoku University.  S.V.F.~was supported by
the National Science Foundation under Grant No.~DMR-0239109,
and E.~R.~W.~was supported by the NSF under Grant
No.~CHE-0910707.  We thank A.~Schofield and W.~C.~K.~Poon for
providing our colloidal samples.

\footnotesize{
\bibliography{eric}

\providecommand*{\mcitethebibliography}{\thebibliography}
\csname @ifundefined\endcsname{endmcitethebibliography}
{\let\endmcitethebibliography\endthebibliography}{}
\begin{mcitethebibliography}{67}
\providecommand*{\natexlab}[1]{#1}
\providecommand*{\mciteSetBstSublistMode}[1]{}
\providecommand*{\mciteSetBstMaxWidthForm}[2]{}
\providecommand*{\mciteBstWouldAddEndPuncttrue}
  {\def\EndOfBibitem{\unskip.}}
\providecommand*{\mciteBstWouldAddEndPunctfalse}
  {\let\EndOfBibitem\relax}
\providecommand*{\mciteSetBstMidEndSepPunct}[3]{}
\providecommand*{\mciteSetBstSublistLabelBeginEnd}[3]{}
\providecommand*{\EndOfBibitem}{}
\mciteSetBstSublistMode{f}
\mciteSetBstMaxWidthForm{subitem}
{(\emph{\alph{mcitesubitemcount}})}
\mciteSetBstSublistLabelBeginEnd{\mcitemaxwidthsubitemform\space}
{\relax}{\relax}

\bibitem[Angell(1995)]{angell95}
C.~A. Angell, \emph{Science}, 1995, \textbf{267}, 1924--1935\relax
\mciteBstWouldAddEndPuncttrue
\mciteSetBstMidEndSepPunct{\mcitedefaultmidpunct}
{\mcitedefaultendpunct}{\mcitedefaultseppunct}\relax
\EndOfBibitem
\bibitem[Stillinger(1995)]{stillinger95}
F.~H. Stillinger, \emph{Science}, 1995, \textbf{267}, 1935--1939\relax
\mciteBstWouldAddEndPuncttrue
\mciteSetBstMidEndSepPunct{\mcitedefaultmidpunct}
{\mcitedefaultendpunct}{\mcitedefaultseppunct}\relax
\EndOfBibitem
\bibitem[Angell \emph{et~al.}(2000)Angell, Ngai, McKenna, McMillan, and
  Martin]{angell00}
C.~A. Angell, K.~L. Ngai, G.~B. McKenna, P.~F. McMillan and S.~W. Martin,
  \emph{J. App. Phys.}, 2000, \textbf{88}, 3113--3157\relax
\mciteBstWouldAddEndPuncttrue
\mciteSetBstMidEndSepPunct{\mcitedefaultmidpunct}
{\mcitedefaultendpunct}{\mcitedefaultseppunct}\relax
\EndOfBibitem
\bibitem[Adam and Gibbs(1965)]{gibbs65}
G.~Adam and J.~H. Gibbs, \emph{J. Chem. Phys.}, 1965, \textbf{43},
  139--146\relax
\mciteBstWouldAddEndPuncttrue
\mciteSetBstMidEndSepPunct{\mcitedefaultmidpunct}
{\mcitedefaultendpunct}{\mcitedefaultseppunct}\relax
\EndOfBibitem
\bibitem[Ediger(2000)]{ediger00}
M.~D. Ediger, \emph{Annu. Rev. Phys. Chem.}, 2000, \textbf{51}, 99--128\relax
\mciteBstWouldAddEndPuncttrue
\mciteSetBstMidEndSepPunct{\mcitedefaultmidpunct}
{\mcitedefaultendpunct}{\mcitedefaultseppunct}\relax
\EndOfBibitem
\bibitem[Donati \emph{et~al.}(1998)Donati, Douglas, Kob, Plimpton, Poole, and
  Glotzer]{donati98}
C.~Donati, J.~F. Douglas, W.~Kob, S.~J. Plimpton, P.~H. Poole and S.~C.
  Glotzer, \emph{Phys. Rev. Lett.}, 1998, \textbf{80}, 2338--2341\relax
\mciteBstWouldAddEndPuncttrue
\mciteSetBstMidEndSepPunct{\mcitedefaultmidpunct}
{\mcitedefaultendpunct}{\mcitedefaultseppunct}\relax
\EndOfBibitem
\bibitem[Pusey and van Megen(1986)]{pusey86}
P.~N. Pusey and W.~van Megen, \emph{Nature}, 1986, \textbf{320}, 340--342\relax
\mciteBstWouldAddEndPuncttrue
\mciteSetBstMidEndSepPunct{\mcitedefaultmidpunct}
{\mcitedefaultendpunct}{\mcitedefaultseppunct}\relax
\EndOfBibitem
\bibitem[van Megen and Pusey(1991)]{vanmegen91}
W.~van Megen and P.~N. Pusey, \emph{Phys. Rev. A}, 1991, \textbf{43},
  5429--5441\relax
\mciteBstWouldAddEndPuncttrue
\mciteSetBstMidEndSepPunct{\mcitedefaultmidpunct}
{\mcitedefaultendpunct}{\mcitedefaultseppunct}\relax
\EndOfBibitem
\bibitem[Segr\`{e} \emph{et~al.}(1995)Segr\`{e}, Meeker, Pusey, and
  Poon]{segre95}
P.~N. Segr\`{e}, S.~P. Meeker, P.~N. Pusey and W.~C.~K. Poon, \emph{Phys. Rev.
  Lett.}, 1995, \textbf{75}, 958--961\relax
\mciteBstWouldAddEndPuncttrue
\mciteSetBstMidEndSepPunct{\mcitedefaultmidpunct}
{\mcitedefaultendpunct}{\mcitedefaultseppunct}\relax
\EndOfBibitem
\bibitem[Cheng \emph{et~al.}(2002)Cheng, Zhu, Chaikin, Phan, and
  Russel]{cheng02}
Z.~Cheng, J.~Zhu, P.~M. Chaikin, S.-E. Phan and W.~B. Russel, \emph{Phys. Rev.
  E}, 2002, \textbf{65}, 041405\relax
\mciteBstWouldAddEndPuncttrue
\mciteSetBstMidEndSepPunct{\mcitedefaultmidpunct}
{\mcitedefaultendpunct}{\mcitedefaultseppunct}\relax
\EndOfBibitem
\bibitem[Mason and Weitz(1995)]{mason95glass}
T.~G. Mason and D.~A. Weitz, \emph{Phys. Rev. Lett.}, 1995, \textbf{75},
  2770--2773\relax
\mciteBstWouldAddEndPuncttrue
\mciteSetBstMidEndSepPunct{\mcitedefaultmidpunct}
{\mcitedefaultendpunct}{\mcitedefaultseppunct}\relax
\EndOfBibitem
\bibitem[Bartsch(1995)]{bartsch95}
E.~Bartsch, \emph{J. Non-Cryst. Solids}, 1995, \textbf{192-193}, 384--392\relax
\mciteBstWouldAddEndPuncttrue
\mciteSetBstMidEndSepPunct{\mcitedefaultmidpunct}
{\mcitedefaultendpunct}{\mcitedefaultseppunct}\relax
\EndOfBibitem
\bibitem[van Megen \emph{et~al.}(1998)van Megen, Mortensen, Williams, and
  M\"{u}ller]{vanmegen98}
W.~van Megen, T.~C. Mortensen, S.~R. Williams and J.~M\"{u}ller, \emph{Phys.
  Rev. E}, 1998, \textbf{58}, 6073--6085\relax
\mciteBstWouldAddEndPuncttrue
\mciteSetBstMidEndSepPunct{\mcitedefaultmidpunct}
{\mcitedefaultendpunct}{\mcitedefaultseppunct}\relax
\EndOfBibitem
\bibitem[Brambilla \emph{et~al.}(2009)Brambilla, El~Masri, Pierno, Berthier,
  Cipelletti, Petekidis, and Schofield]{brambilla09}
G.~Brambilla, D.~E.~M. El~Masri, M.~Pierno, L.~Berthier, L.~Cipelletti,
  G.~Petekidis and A.~B. Schofield, \emph{Phys. Rev. Lett.}, 2009,
  \textbf{102}, 085703\relax
\mciteBstWouldAddEndPuncttrue
\mciteSetBstMidEndSepPunct{\mcitedefaultmidpunct}
{\mcitedefaultendpunct}{\mcitedefaultseppunct}\relax
\EndOfBibitem
\bibitem[van Megen and Williams(2010)]{vanmegen10}
W.~van Megen and S.~R. Williams, \emph{Phys. Rev. Lett.}, 2010, \textbf{104},
  169601\relax
\mciteBstWouldAddEndPuncttrue
\mciteSetBstMidEndSepPunct{\mcitedefaultmidpunct}
{\mcitedefaultendpunct}{\mcitedefaultseppunct}\relax
\EndOfBibitem
\bibitem[Brambilla \emph{et~al.}(2010)Brambilla, Masri, Pierno, Berthier,
  Cipelletti, Petekidis, and Schofield]{brambilla10}
G.~Brambilla, D.~E. Masri, M.~Pierno, L.~Berthier, L.~Cipelletti, G.~Petekidis
  and A.~Schofield, \emph{Phys. Rev. Lett.}, 2010, \textbf{104}, 169602\relax
\mciteBstWouldAddEndPuncttrue
\mciteSetBstMidEndSepPunct{\mcitedefaultmidpunct}
{\mcitedefaultendpunct}{\mcitedefaultseppunct}\relax
\EndOfBibitem
\bibitem[van Blaaderen and Wiltzius(1995)]{vanblaaderen95}
A.~van Blaaderen and P.~Wiltzius, \emph{Science}, 1995, \textbf{270},
  1177--1179\relax
\mciteBstWouldAddEndPuncttrue
\mciteSetBstMidEndSepPunct{\mcitedefaultmidpunct}
{\mcitedefaultendpunct}{\mcitedefaultseppunct}\relax
\EndOfBibitem
\bibitem[Marcus \emph{et~al.}(1999)Marcus, Schofield, and Rice]{marcus99}
A.~H. Marcus, J.~Schofield and S.~A. Rice, \emph{Phys. Rev. E}, 1999,
  \textbf{60}, 5725--5736\relax
\mciteBstWouldAddEndPuncttrue
\mciteSetBstMidEndSepPunct{\mcitedefaultmidpunct}
{\mcitedefaultendpunct}{\mcitedefaultseppunct}\relax
\EndOfBibitem
\bibitem[Kegel and van Blaaderen(2000)]{kegel00}
W.~K. Kegel and A.~van Blaaderen, \emph{Science}, 2000, \textbf{287},
  290--293\relax
\mciteBstWouldAddEndPuncttrue
\mciteSetBstMidEndSepPunct{\mcitedefaultmidpunct}
{\mcitedefaultendpunct}{\mcitedefaultseppunct}\relax
\EndOfBibitem
\bibitem[Weeks \emph{et~al.}(2000)Weeks, Crocker, Levitt, Schofield, and
  Weitz]{weeks00}
E.~R. Weeks, J.~C. Crocker, A.~C. Levitt, A.~Schofield and D.~A. Weitz,
  \emph{Science}, 2000, \textbf{287}, 627--631\relax
\mciteBstWouldAddEndPuncttrue
\mciteSetBstMidEndSepPunct{\mcitedefaultmidpunct}
{\mcitedefaultendpunct}{\mcitedefaultseppunct}\relax
\EndOfBibitem
\bibitem[K\"{o}nig \emph{et~al.}(2005)K\"{o}nig, Hund, Zahn, and
  Maret]{konig05}
H.~K\"{o}nig, R.~Hund, K.~Zahn and G.~Maret, \emph{Euro. Phys. J. E}, 2005,
  \textbf{18}, 287--293\relax
\mciteBstWouldAddEndPuncttrue
\mciteSetBstMidEndSepPunct{\mcitedefaultmidpunct}
{\mcitedefaultendpunct}{\mcitedefaultseppunct}\relax
\EndOfBibitem
\bibitem[Courtland and Weeks(2003)]{courtland03}
R.~E. Courtland and E.~R. Weeks, \emph{J. Phys.: Cond. Matt.}, 2003,
  \textbf{15}, S359--S365\relax
\mciteBstWouldAddEndPuncttrue
\mciteSetBstMidEndSepPunct{\mcitedefaultmidpunct}
{\mcitedefaultendpunct}{\mcitedefaultseppunct}\relax
\EndOfBibitem
\bibitem[Cipelletti \emph{et~al.}(2003)Cipelletti, Bissig, Trappe, Ballesta,
  and Mazoyer]{cipelletti03}
L.~Cipelletti, H.~Bissig, V.~Trappe, P.~Ballesta and S.~Mazoyer, \emph{J.
  Phys.: Condens. Matter}, 2003, \textbf{15}, S257--S262\relax
\mciteBstWouldAddEndPuncttrue
\mciteSetBstMidEndSepPunct{\mcitedefaultmidpunct}
{\mcitedefaultendpunct}{\mcitedefaultseppunct}\relax
\EndOfBibitem
\bibitem[Simeonova and Kegel(2004)]{kegel04}
N.~B. Simeonova and W.~K. Kegel, \emph{Phys. Rev. Lett.}, 2004, \textbf{93},
  035701\relax
\mciteBstWouldAddEndPuncttrue
\mciteSetBstMidEndSepPunct{\mcitedefaultmidpunct}
{\mcitedefaultendpunct}{\mcitedefaultseppunct}\relax
\EndOfBibitem
\bibitem[El~Masri \emph{et~al.}(2005)El~Masri, Pierno, Berthier, and
  Cipelletti]{cipelletti05}
D.~El~Masri, M.~Pierno, L.~Berthier and L.~Cipelletti, \emph{J. Phys.: Cond.
  Matt.}, 2005, \textbf{17}, S3543--S3549\relax
\mciteBstWouldAddEndPuncttrue
\mciteSetBstMidEndSepPunct{\mcitedefaultmidpunct}
{\mcitedefaultendpunct}{\mcitedefaultseppunct}\relax
\EndOfBibitem
\bibitem[Cianci \emph{et~al.}(2006)Cianci, Courtland, and Weeks]{cianci06ssc}
G.~C. Cianci, R.~E. Courtland and E.~R. Weeks, \emph{Solid State Comm.}, 2006,
  \textbf{139}, 599--604\relax
\mciteBstWouldAddEndPuncttrue
\mciteSetBstMidEndSepPunct{\mcitedefaultmidpunct}
{\mcitedefaultendpunct}{\mcitedefaultseppunct}\relax
\EndOfBibitem
\bibitem[Yunker \emph{et~al.}(2009)Yunker, Zhang, Aptowicz, and Yodh]{yunker09}
P.~Yunker, Z.~Zhang, K.~Aptowicz and A.~Yodh, \emph{Phys. Rev. Lett.}, 2009,
  \textbf{103}, 115701\relax
\mciteBstWouldAddEndPuncttrue
\mciteSetBstMidEndSepPunct{\mcitedefaultmidpunct}
{\mcitedefaultendpunct}{\mcitedefaultseppunct}\relax
\EndOfBibitem
\bibitem[Nugent \emph{et~al.}(2007)Nugent, Edmond, Patel, and
  Weeks]{nugent07prl}
C.~R. Nugent, K.~V. Edmond, H.~N. Patel and E.~R. Weeks, \emph{Phys. Rev.
  Lett.}, 2007, \textbf{99}, 025702\relax
\mciteBstWouldAddEndPuncttrue
\mciteSetBstMidEndSepPunct{\mcitedefaultmidpunct}
{\mcitedefaultendpunct}{\mcitedefaultseppunct}\relax
\EndOfBibitem
\bibitem[Sarangapani and Zhu(2008)]{sarangapani08}
P.~S. Sarangapani and Y.~Zhu, \emph{Phys. Rev. E}, 2008, \textbf{77},
  010501\relax
\mciteBstWouldAddEndPuncttrue
\mciteSetBstMidEndSepPunct{\mcitedefaultmidpunct}
{\mcitedefaultendpunct}{\mcitedefaultseppunct}\relax
\EndOfBibitem
\bibitem[Conrad \emph{et~al.}(2006)Conrad, Dhillon, Weeks, Reichman, and
  Weitz]{conrad06}
J.~C. Conrad, P.~P. Dhillon, E.~R. Weeks, D.~R. Reichman and D.~A. Weitz,
  \emph{Phys. Rev. Lett.}, 2006, \textbf{97}, 265701\relax
\mciteBstWouldAddEndPuncttrue
\mciteSetBstMidEndSepPunct{\mcitedefaultmidpunct}
{\mcitedefaultendpunct}{\mcitedefaultseppunct}\relax
\EndOfBibitem
\bibitem[Zhu \emph{et~al.}(1997)Zhu, Li, Rogers, Meyer, Ottewill, Sts-73,
  Russel, and Chaikin]{zhu97}
J.~Zhu, M.~Li, R.~Rogers, W.~Meyer, R.~H. Ottewill, Sts-73, W.~B. Russel and
  P.~M. Chaikin, \emph{Nature}, 1997, \textbf{387}, 883--885\relax
\mciteBstWouldAddEndPuncttrue
\mciteSetBstMidEndSepPunct{\mcitedefaultmidpunct}
{\mcitedefaultendpunct}{\mcitedefaultseppunct}\relax
\EndOfBibitem
\bibitem[Gasser \emph{et~al.}(2001)Gasser, Weeks, Schofield, Pusey, and
  Weitz]{gasser01}
U.~Gasser, E.~R. Weeks, A.~Schofield, P.~N. Pusey and D.~A. Weitz,
  \emph{Science}, 2001, \textbf{292}, 258--262\relax
\mciteBstWouldAddEndPuncttrue
\mciteSetBstMidEndSepPunct{\mcitedefaultmidpunct}
{\mcitedefaultendpunct}{\mcitedefaultseppunct}\relax
\EndOfBibitem
\bibitem[Lynch \emph{et~al.}(2008)Lynch, Cianci, and Weeks]{lynch08}
J.~M. Lynch, G.~C. Cianci and E.~R. Weeks, \emph{Phys. Rev. E}, 2008,
  \textbf{78}, 031410\relax
\mciteBstWouldAddEndPuncttrue
\mciteSetBstMidEndSepPunct{\mcitedefaultmidpunct}
{\mcitedefaultendpunct}{\mcitedefaultseppunct}\relax
\EndOfBibitem
\bibitem[Weeks and Weitz(2002)]{weeks02}
E.~R. Weeks and D.~A. Weitz, \emph{Phys. Rev. Lett.}, 2002, \textbf{89},
  095704\relax
\mciteBstWouldAddEndPuncttrue
\mciteSetBstMidEndSepPunct{\mcitedefaultmidpunct}
{\mcitedefaultendpunct}{\mcitedefaultseppunct}\relax
\EndOfBibitem
\bibitem[Conrad \emph{et~al.}(2005)Conrad, Starr, and Weitz]{conrad05}
J.~C. Conrad, F.~W. Starr and D.~A. Weitz, \emph{J. Phys. Chem. B}, 2005,
  \textbf{109}, 21235--21240\relax
\mciteBstWouldAddEndPuncttrue
\mciteSetBstMidEndSepPunct{\mcitedefaultmidpunct}
{\mcitedefaultendpunct}{\mcitedefaultseppunct}\relax
\EndOfBibitem
\bibitem[Widmer-Cooper \emph{et~al.}(2004)Widmer-Cooper, Harrowell, and
  Fynewever]{harrowell04}
A.~Widmer-Cooper, P.~Harrowell and H.~Fynewever, \emph{Phys. Rev. Lett.}, 2004,
  \textbf{93}, 135701\relax
\mciteBstWouldAddEndPuncttrue
\mciteSetBstMidEndSepPunct{\mcitedefaultmidpunct}
{\mcitedefaultendpunct}{\mcitedefaultseppunct}\relax
\EndOfBibitem
\bibitem[Widmer-Cooper and Harrowell(2005)]{widmercooper05}
A.~Widmer-Cooper and P.~Harrowell, \emph{J. Phys.: Cond. Matt.}, 2005,
  \textbf{17}, S4025--S4034\relax
\mciteBstWouldAddEndPuncttrue
\mciteSetBstMidEndSepPunct{\mcitedefaultmidpunct}
{\mcitedefaultendpunct}{\mcitedefaultseppunct}\relax
\EndOfBibitem
\bibitem[Prasad \emph{et~al.}(2007)Prasad, Semwogerere, and Weeks]{prasad07}
V.~Prasad, D.~Semwogerere and E.~R. Weeks, \emph{J. Phys.: Cond. Matt.}, 2007,
  \textbf{19}, 113102\relax
\mciteBstWouldAddEndPuncttrue
\mciteSetBstMidEndSepPunct{\mcitedefaultmidpunct}
{\mcitedefaultendpunct}{\mcitedefaultseppunct}\relax
\EndOfBibitem
\bibitem[Poole \emph{et~al.}(1998)Poole, Donati, and Glotzer]{poole98}
P.~H. Poole, C.~Donati and S.~C. Glotzer, \emph{Physica A: Statistical and
  Theoretical Physics}, 1998, \textbf{261}, 51--59\relax
\mciteBstWouldAddEndPuncttrue
\mciteSetBstMidEndSepPunct{\mcitedefaultmidpunct}
{\mcitedefaultendpunct}{\mcitedefaultseppunct}\relax
\EndOfBibitem
\bibitem[Donati \emph{et~al.}(1999)Donati, Glotzer, and Poole]{donati99}
C.~Donati, S.~C. Glotzer and P.~H. Poole, \emph{Phys. Rev. Lett.}, 1999,
  \textbf{82}, 5064--5067\relax
\mciteBstWouldAddEndPuncttrue
\mciteSetBstMidEndSepPunct{\mcitedefaultmidpunct}
{\mcitedefaultendpunct}{\mcitedefaultseppunct}\relax
\EndOfBibitem
\bibitem[Doliwa and Heuer(2000)]{doliwa00}
B.~Doliwa and A.~Heuer, \emph{Phys. Rev. E}, 2000, \textbf{61},
  6898--6908\relax
\mciteBstWouldAddEndPuncttrue
\mciteSetBstMidEndSepPunct{\mcitedefaultmidpunct}
{\mcitedefaultendpunct}{\mcitedefaultseppunct}\relax
\EndOfBibitem
\bibitem[Weeks \emph{et~al.}(2007)Weeks, Crocker, and Weitz]{weeks07cor}
E.~R. Weeks, J.~C. Crocker and D.~A. Weitz, \emph{J. Phys.: Cond. Matt.}, 2007,
  \textbf{19}, 205131\relax
\mciteBstWouldAddEndPuncttrue
\mciteSetBstMidEndSepPunct{\mcitedefaultmidpunct}
{\mcitedefaultendpunct}{\mcitedefaultseppunct}\relax
\EndOfBibitem
\bibitem[Glotzer \emph{et~al.}(2000)Glotzer, Novikov, and
  Schr{\o}~der]{glotzer00_2}
S.~C. Glotzer, V.~N. Novikov and T.~B. Schr{\o}~der, \emph{J. Chem. Phys.},
  2000, \textbf{112}, 509--512\relax
\mciteBstWouldAddEndPuncttrue
\mciteSetBstMidEndSepPunct{\mcitedefaultmidpunct}
{\mcitedefaultendpunct}{\mcitedefaultseppunct}\relax
\EndOfBibitem
\bibitem[Lacevic \emph{et~al.}(2003)Lacevic, Starr, Schroder, and
  Glotzer]{glotzer03a}
N.~Lacevic, F.~W. Starr, T.~B. Schroder and S.~C. Glotzer, \emph{J. Chem.
  Phys.}, 2003, \textbf{119}, 7372--7387\relax
\mciteBstWouldAddEndPuncttrue
\mciteSetBstMidEndSepPunct{\mcitedefaultmidpunct}
{\mcitedefaultendpunct}{\mcitedefaultseppunct}\relax
\EndOfBibitem
\bibitem[Keys \emph{et~al.}(2007)Keys, Abate, Glotzer, and Durian]{keys07}
A.~S. Keys, A.~R. Abate, S.~C. Glotzer and D.~J. Durian, \emph{Nature Physics},
  2007, \textbf{3}, 260--264\relax
\mciteBstWouldAddEndPuncttrue
\mciteSetBstMidEndSepPunct{\mcitedefaultmidpunct}
{\mcitedefaultendpunct}{\mcitedefaultseppunct}\relax
\EndOfBibitem
\bibitem[Dinsmore \emph{et~al.}(2001)Dinsmore, Weeks, Prasad, Levitt, and
  Weitz]{dinsmore01}
A.~D. Dinsmore, E.~R. Weeks, V.~Prasad, A.~C. Levitt and D.~A. Weitz,
  \emph{App. Optics}, 2001, \textbf{40}, 4152--4159\relax
\mciteBstWouldAddEndPuncttrue
\mciteSetBstMidEndSepPunct{\mcitedefaultmidpunct}
{\mcitedefaultendpunct}{\mcitedefaultseppunct}\relax
\EndOfBibitem
\bibitem[Crocker and Grier(1996)]{crocker96}
J.~C. Crocker and D.~G. Grier, \emph{J. Colloid Interf. Sci.}, 1996,
  \textbf{179}, 298--310\relax
\mciteBstWouldAddEndPuncttrue
\mciteSetBstMidEndSepPunct{\mcitedefaultmidpunct}
{\mcitedefaultendpunct}{\mcitedefaultseppunct}\relax
\EndOfBibitem
\bibitem[Einstein(1905)]{einstein1905a}
A.~Einstein, \emph{Annalen der Physik (Leipzig)}, 1905, \textbf{17},
  549--560\relax
\mciteBstWouldAddEndPuncttrue
\mciteSetBstMidEndSepPunct{\mcitedefaultmidpunct}
{\mcitedefaultendpunct}{\mcitedefaultseppunct}\relax
\EndOfBibitem
\bibitem[Sutherland(1905)]{sutherland1905}
W.~Sutherland, \emph{Phil. Mag.}, 1905, \textbf{9}, 781--785\relax
\mciteBstWouldAddEndPuncttrue
\mciteSetBstMidEndSepPunct{\mcitedefaultmidpunct}
{\mcitedefaultendpunct}{\mcitedefaultseppunct}\relax
\EndOfBibitem
\bibitem[Rabani \emph{et~al.}(1997)Rabani, Gezelter, and Berne]{berne97}
E.~Rabani, D.~J. Gezelter and B.~J. Berne, \emph{J. Chem. Phys.}, 1997,
  \textbf{107}, 6867--6876\relax
\mciteBstWouldAddEndPuncttrue
\mciteSetBstMidEndSepPunct{\mcitedefaultmidpunct}
{\mcitedefaultendpunct}{\mcitedefaultseppunct}\relax
\EndOfBibitem
\bibitem[Doliwa and Heuer(1998)]{doliwa98}
B.~Doliwa and A.~Heuer, \emph{Phys. Rev. Lett.}, 1998, \textbf{80},
  4915--4918\relax
\mciteBstWouldAddEndPuncttrue
\mciteSetBstMidEndSepPunct{\mcitedefaultmidpunct}
{\mcitedefaultendpunct}{\mcitedefaultseppunct}\relax
\EndOfBibitem
\bibitem[Kasper \emph{et~al.}(1998)Kasper, Bartsch, and Sillescu]{bartsch98}
A.~Kasper, E.~Bartsch and H.~Sillescu, \emph{Langmuir}, 1998, \textbf{14},
  5004--5010\relax
\mciteBstWouldAddEndPuncttrue
\mciteSetBstMidEndSepPunct{\mcitedefaultmidpunct}
{\mcitedefaultendpunct}{\mcitedefaultseppunct}\relax
\EndOfBibitem
\bibitem[Weeks and Weitz(2002)]{weeks02sub}
E.~R. Weeks and D.~A. Weitz, \emph{Chem. Phys.}, 2002, \textbf{284},
  361--367\relax
\mciteBstWouldAddEndPuncttrue
\mciteSetBstMidEndSepPunct{\mcitedefaultmidpunct}
{\mcitedefaultendpunct}{\mcitedefaultseppunct}\relax
\EndOfBibitem
\bibitem[Reis \emph{et~al.}(2007)Reis, Ingale, and Shattuck]{shattuck07}
P.~M. Reis, R.~A. Ingale and M.~D. Shattuck, \emph{Phys. Rev. Lett.}, 2007,
  \textbf{98}, 188301\relax
\mciteBstWouldAddEndPuncttrue
\mciteSetBstMidEndSepPunct{\mcitedefaultmidpunct}
{\mcitedefaultendpunct}{\mcitedefaultseppunct}\relax
\EndOfBibitem
\bibitem[Ngai and Rendell(1998)]{ngai98}
K.~L. Ngai and R.~W. Rendell, \emph{Phil. Mag. B}, 1998, \textbf{77},
  621--631\relax
\mciteBstWouldAddEndPuncttrue
\mciteSetBstMidEndSepPunct{\mcitedefaultmidpunct}
{\mcitedefaultendpunct}{\mcitedefaultseppunct}\relax
\EndOfBibitem
\bibitem[van Megen \emph{et~al.}(2009)van Megen, Martinez, and
  Bryant]{vanmegen09}
W.~van Megen, V.~A. Martinez and G.~Bryant, \emph{Phys. Rev. Lett.}, 2009,
  \textbf{102}, 168301\relax
\mciteBstWouldAddEndPuncttrue
\mciteSetBstMidEndSepPunct{\mcitedefaultmidpunct}
{\mcitedefaultendpunct}{\mcitedefaultseppunct}\relax
\EndOfBibitem
\bibitem[Kob \emph{et~al.}(1997)Kob, Donati, Plimpton, Poole, and
  Glotzer]{kob97}
W.~Kob, C.~Donati, S.~J. Plimpton, P.~H. Poole and S.~C. Glotzer, \emph{Phys.
  Rev. Lett.}, 1997, \textbf{79}, 2827--2830\relax
\mciteBstWouldAddEndPuncttrue
\mciteSetBstMidEndSepPunct{\mcitedefaultmidpunct}
{\mcitedefaultendpunct}{\mcitedefaultseppunct}\relax
\EndOfBibitem
\bibitem[Rahman(1964)]{rahman64}
A.~Rahman, \emph{Phys. Rev.}, 1964, \textbf{136}, A405--A411\relax
\mciteBstWouldAddEndPuncttrue
\mciteSetBstMidEndSepPunct{\mcitedefaultmidpunct}
{\mcitedefaultendpunct}{\mcitedefaultseppunct}\relax
\EndOfBibitem
\bibitem[Hoffman(1992)]{hoffman92}
R.~L. Hoffman, \emph{J. Rheology}, 1992, \textbf{36}, 947--965\relax
\mciteBstWouldAddEndPuncttrue
\mciteSetBstMidEndSepPunct{\mcitedefaultmidpunct}
{\mcitedefaultendpunct}{\mcitedefaultseppunct}\relax
\EndOfBibitem
\bibitem[D'Haene and Mewis(1994)]{mewis94}
P.~D'Haene and J.~Mewis, \emph{Rheologica Acta}, 1994, \textbf{33},
  165--174\relax
\mciteBstWouldAddEndPuncttrue
\mciteSetBstMidEndSepPunct{\mcitedefaultmidpunct}
{\mcitedefaultendpunct}{\mcitedefaultseppunct}\relax
\EndOfBibitem
\bibitem[Williams and van Megen(2001)]{vanmegen01}
S.~R. Williams and W.~van Megen, \emph{Phys. Rev. E}, 2001, \textbf{64},
  041502\relax
\mciteBstWouldAddEndPuncttrue
\mciteSetBstMidEndSepPunct{\mcitedefaultmidpunct}
{\mcitedefaultendpunct}{\mcitedefaultseppunct}\relax
\EndOfBibitem
\bibitem[Kawasaki \emph{et~al.}(2007)Kawasaki, Araki, and Tanaka]{kawasaki07}
T.~Kawasaki, T.~Araki and H.~Tanaka, \emph{Phys. Rev. Lett.}, 2007,
  \textbf{99}, 215701\relax
\mciteBstWouldAddEndPuncttrue
\mciteSetBstMidEndSepPunct{\mcitedefaultmidpunct}
{\mcitedefaultendpunct}{\mcitedefaultseppunct}\relax
\EndOfBibitem
\bibitem[Narumi and Tokuyama(2008)]{narumi08}
T.~Narumi and M.~Tokuyama, \emph{Phil. Mag.}, 2008, \textbf{88},
  4169--4175\relax
\mciteBstWouldAddEndPuncttrue
\mciteSetBstMidEndSepPunct{\mcitedefaultmidpunct}
{\mcitedefaultendpunct}{\mcitedefaultseppunct}\relax
\EndOfBibitem
\bibitem[Sarangapani \emph{et~al.}(2008)Sarangapani, Zhao, and
  Zhu]{sarangapani08b}
P.~Sarangapani, J.~Zhao and Y.~Zhu, \emph{J. Chem. Phys.}, 2008, \textbf{129},
  104514\relax
\mciteBstWouldAddEndPuncttrue
\mciteSetBstMidEndSepPunct{\mcitedefaultmidpunct}
{\mcitedefaultendpunct}{\mcitedefaultseppunct}\relax
\EndOfBibitem
\bibitem[G\"{o}tze(1999)]{Gotze99}
W.~G\"{o}tze, \emph{J. Phys.: Cond. Matt.}, 1999, \textbf{11}, A1--A45\relax
\mciteBstWouldAddEndPuncttrue
\mciteSetBstMidEndSepPunct{\mcitedefaultmidpunct}
{\mcitedefaultendpunct}{\mcitedefaultseppunct}\relax
\EndOfBibitem
\bibitem[Berthier \emph{et~al.}(2007)Berthier, Biroli, Bouchaud, Kob, Miyazaki,
  and Reichman]{berthier07a}
L.~Berthier, G.~Biroli, J.~P. Bouchaud, W.~Kob, K.~Miyazaki and D.~R. Reichman,
  \emph{J. Chem. Phys.}, 2007, \textbf{126}, 184503\relax
\mciteBstWouldAddEndPuncttrue
\mciteSetBstMidEndSepPunct{\mcitedefaultmidpunct}
{\mcitedefaultendpunct}{\mcitedefaultseppunct}\relax
\EndOfBibitem
\bibitem[Berthier \emph{et~al.}(2007)Berthier, Biroli, Bouchaud, Kob, Miyazaki,
  and Reichman]{berthier07b}
L.~Berthier, G.~Biroli, J.~P. Bouchaud, W.~Kob, K.~Miyazaki and D.~R. Reichman,
  \emph{J. Chem. Phys.}, 2007, \textbf{126}, 184504\relax
\mciteBstWouldAddEndPuncttrue
\mciteSetBstMidEndSepPunct{\mcitedefaultmidpunct}
{\mcitedefaultendpunct}{\mcitedefaultseppunct}\relax
\EndOfBibitem
\end{mcitethebibliography}
\bibliographystyle{rsc}
}

\end{document}